\documentclass[12pt]{article}
\usepackage{amsthm,amsmath,amsfonts,amssymb,cases}
\usepackage{calc} %für eckiges Integral
\usepackage{comment}
\usepackage{enumerate}
\usepackage{tikz}
\usepackage{pgfplots}
\usepackage[shortlabels]{enumitem}
\usepackage{hyperref}
\usepackage{cleveref}
\usepackage{float}
\usepackage{cite}

\newcommand{\R}{{\mathord{\mathbb R}}}
\newcommand{\Z}{{\mathord{\mathbb Z}}}
\newcommand{\N}{{\mathord{\mathbb N}}}
\newcommand{\C}{{\mathord{\mathbb C}}}

\newcommand{\hh}{\mathfrak{h}}

\newcommand{\muun}[1]{\begin{align} #1 \end{align}}

\renewcommand{\epsilon}{\varepsilon}

\newcommand{\vertiii}[1]{{\left\vert\kern-0.25ex\left\vert\kern-0.25ex\left\vert #1 
    \right\vert\kern-0.25ex\right\vert\kern-0.25ex\right\vert}}

\newcommand{\ben}{\begin{displaymath}}
\newcommand{\een}{\end{displaymath}}
\newcommand{\beqn}{\begin{equation}}
\newcommand{\eeqn}{\end{equation}}
\newcommand{\beqna}{\begin{eqnarray*}}
\newcommand{\eeqna}{\end{eqnarray*}}

\newcommand{\nn}{\nonumber} 

\def\const{{\rm const}\,}

\def\supp{\operatorname{supp}}

\newcommand{\inn}[1]{\langle {#1} \rangle }

\newtheorem{lemma}{Lemma}
\newtheorem{theorem}[lemma]{Theorem}
\newtheorem{remark}[lemma]{Remark}
\newtheorem{proposition}[lemma]{Proposition}

\newtheorem{definition}[lemma]{Definition}
\newtheorem{hyp}{Hypothesis}

\numberwithin{equation}{section}
\numberwithin{lemma}{section}

\makeatletter
\newsavebox{\@brx}
\newcommand{\llangle}[1][]{\savebox{\@brx}{\(\m@th{#1\langle}\)}%
  \mathopen{\copy\@brx\kern-0.5\wd\@brx\usebox{\@brx}}}
\newcommand{\rrangle}[1][]{\savebox{\@brx}{\(\m@th{#1\rangle}\)}%
  \mathclose{\copy\@brx\kern-0.5\wd\@brx\usebox{\@brx}}}
\makeatother

\newlength{\hght}
\newlength{\dpth}

\begin{document}
\title{ Limits  of spectral measures  for  linearly  bounded and for Poisson distributed random potentials}
\author{\vspace{5pt} David Hasler$^1$\footnote{
E-mail: david.hasler@uni-jena.de } \quad  Jannis Koberstein$^1$\footnote{E-mail: jannis.koberstein@eah-jena.de} \\
\vspace{-4pt} \small{$1.$ Department of Mathematics,
Friedrich Schiller  University  Jena} \\ \small{Jena, Germany } }
\date{}
\maketitle

\begin{abstract}
We show the existence of infinite volume  limits of resolvents and spectral measures for 
 a  class of    Schr\"odinger operators with  linearly bounded potentials.  
We then apply this result to Schr\"odinger operators with a Poisson distributed random potential.    
\end{abstract}

\section{Introduction}

Schrödinger operators are a class of partial differential operatators used to describe 
quantum mechanical properties of  physical systems. 

For example, the behavior of a single electron in a solid can be analyzed using a Schrödinger operator, where, in a simplified approximation, the interaction with other particles is captured by a static potential.
In many cases, the potential is subject to randomness due to impurities or irregularities in the solid. This leads to  the concept of random Schrödinger operators, which were first studied by Anderson in his seminal work \cite{Anderson.1958}.

Spectral measures of Schrödinger operators encode important spectral properties. 
 They can be  used to determine  spectral types, such as point 
 spectrum  or absolutely continuous spectrum, which are closely linked 
 to physical phenomena like bound and extend states. 
 These states are key to understanding the dynamics of particles and are
  therefore essential  in the study of properties like electrical conductance.

For random Schrödinger operators, a key quantity is the integrated density of states (IDS), which can be expressed as the expectation of a normalized sum of spectral measures. The IDS plays a central role, as it provides the average number of eigenstates (per unit volume) below a given energy level. 
This quantity is particularly useful for understanding the macroscopic behavior of disordered systems.
 Properties such as the existence of the  IDS for a broad class of random Schrödinger operators have been 
 intensively investigated, see e.g. \cite{KirschMetzger.2007,Wegner.1981,CombesHislopKlopp.2003,CombesHislopKlopp.2007,CombesHislop.1994,Pastur.1973,KirschMartinelli.1982}.
 On the other hand, expectations of powers  of spectral measures can be used 
 to detect  extended states.   In particular in   the small disorder regime, this turns out
 to be   a challenging 
subject, and has beend addressed in several publications cf. \cite{AcostaKlein.1992,Klein.1998,AizenmanSimsWarzel.2006,FroeseHaslerSpitzer.2007,AizenmanWarzel.2013,Lloyd.1969,KirschKrishna.2020,AnantharamIngremeauSabriWinn.2021}.

To compute physical quantities  for  operators defined on the entire infinite space, a common technique is to restrict the operator to a finite box and study how these restrictions behave as the box tends to infinity.
For example   when studying  Poisson distributed random potentials, this is convenient,  because of the unbounded nature of such potentials on infinite space.  

In this paper we consider  Schrödigner operators on
a finite box with periodic boundary conditions.  We show  the  existence of infinite volume limits of resolvents and spectral measures. 
 Our result  covers deterministic    potentials which may grow linearly at infinity.
 This growth  requires some 
effort to  control the spacial decay of resolvents, which we establish by means of 
a Combes-Thomas type argument (see  \cite{CombesThomas.1973,SR4}). 
The aformentioned generality of potentials  allows us to extend our results  to a wide class of Poisson-distributed random potentials, where we demonstrate that the infinite-volume limit of resolvents and  spectral measures exist almost surely. This result serves as a foundation for deriving asymptotic expansions in the coupling strength of spectral measures, which we plan to address in future work, cf. \cite{HaslerKoberstein.2024-1,HaslerKoberstein.2024-2}.

The remainder of this paper is organized as follows. In Section \ref{Model}, we introduce the model of a Schrödinger operator with Poisson-distributed random potential, following a setup similar to \cite{ErdosSalmhoferYau.2008a}. We consider restrictions to boxes $\Lambda_L$ of side length $L > 0$, with periodic boundary conditions. Additionally, we present our main results concerning this model.  In Theorem \ref{feynmankac-3}, we state our first main result, which establishes the existence of the weak limit of resolvents as $L \to \infty$. Theorem \ref{ThmFolgPMTgen-0} provides our second main result, demonstrating the existence of the infinite-volume limit of the corresponding spectral measures in both the vague and weak senses.

Section \ref{proofs} contains the main technical work of this paper. In this section 
we consider a larger class of potentials,  defined by Hypothesis \ref{hyp1}, which satisfy a  linear bound condition.  
In  Theorem \ref{mainThm} we prove a  generalized version of \ref{feynmankac-3}, which holds
for this larger class of potentials. That is we show that resolvents converge weakly as the box size $L$
tends to infinity. For this result we use the   Combes-Thomas type argument to quantify the exponential decay of the resolvent.  
We can then prove  in Section \ref{specmeasgen} in Theorem  \ref{ThmFolgPMTgen}
 the convergence of spectral measures for this larger class of potentials. 
Theorem  \ref{ThmFolgPMTgen} is a generalization  of Theorem \ref{ThmFolgPMTgen-0}.
Its proof makes use of the  Portmanteau Theorem, cf. Theorem \ref{smallPMT}, given in the appendix.

Finally,  the appendix consists of three parts. Part \ref{ApA} provides probabilistic estimates on the random potential. Part \ref{ApB} offers estimates on the operator norm of the resolvent and methods for approximating sums with integrals.  Part \ref{ApC} introduces the concepts of vague and weak convergence of measure sequences and includes the version of the Portmanteau theorem \ref{smallPMT}, which we apply in this paper.

\section{Poisson distributed random potential: model and results} 
\label{Model}
In this section we  introduce  a class of Schroedinger operators with Poisson distributed random
potential, and state the main results about  this model concerning  convergence of resolvents and spectral measures.
We note that the definition of the model follows  the one given in 
\cite[Sections 2.1, 3.3]{ErdosSalmhoferYau.2008a} closely.

We define the d-dimensional finite box  $\Lambda_L := \left[ - \frac{1}{2} L , \frac{1}{2} L \right)^d \subset \R^d$ of size $L>0$.
%In this way, for technical convenience, we avoid the infinite summation in \eqref{esy:2.3}. 
For $f,  g \in L^2(\Lambda_L)$ let $$ \inn{ f  ,  g }_L := \int_{\Lambda_L}  \overline{f(x)} g(x) dx , \quad \| f \|_L := \left( \int_{\Lambda_L}  |f(x)|^2  dx \right)^{1/2}  $$ 
%$ \inn{ \cdot , \cdot }_{L^2(\Lambda_L)}$ and $\| \cdot \|_{L^2(\Lambda_L)}$ denote the canonical scalar product and the norm of the Hilber space  $L^2(\Lambda_L)$. 
If it is clear from the 
context what the inner product or  the norm is,   we shall occasionally drop the subscript $L$.
 The kinetic energy will be given by the periodic Laplacian. To introduce this operator it is convenient to 
work in terms of Fourier series. Let  
$\Lambda_L^*  = ( \frac{1}{L} \Z  )^d=\{ (k_1 ,  \ldots ,  k_d) : \forall j=1, \ldots , d,  \exists m_j \in \Z,: k_j =\frac{m_j}{L} \}$ denote  
the so-called dual lattice. We introduce the notation 
\begin{equation}
%\tcr{\sqrint f(p) dp :=} \int_{*} f(p) dp  := 
\int_{\Lambda_L^*} f(p) dp  := \frac{1}{|\Lambda_L |} \sum_{p \in \Lambda_L^*} f(p) , 
\end{equation} 
where $|\Lambda_L | := L^d$ is the volume of the box. 
The sum  $\int_{\Lambda_L^*}f(p)dp$ can be interpreted as a Riemann-sum, which  converges to the integral $\int f(p) dp $ 
as   $L \to \infty$, provided $f$ has  sufficient  decay at infinity and sufficient regularity.  
For any $f \in L^1(\Lambda_L)$ we define the Fourier series 
\begin{align} 
\label{FToff} 
\hat{f}(p) = \int_{\Lambda_L} e^{ - 2 \pi i p \cdot x}  f(x) dx  \quad  \text{ for } p \in \Lambda_L^*  . 
\end{align} 
%Note that the Fourier  $f \in L^2(\Lambda_L)$, then  $\hat{f} \in \ell^2(\Lambda_L^*)$.
  For $g    \in \ell^1(\Lambda_L^*)$ we define 
$$
\check{g}(x) = \int_{\Lambda^*_L} e^{  2 \pi i  p \cdot x } g(p) dp  = \frac{1}{|\Lambda_L |} \sum_{p \in \Lambda_L^*} e^{  2 \pi i  p \cdot x } g(p) \text{ for }  x \in \Lambda_L. 
$$
Note that 
$\check{(\cdot)}$ extends uniquely to a continuous linear map  on $ \ell^2(\Lambda_L^*)$.  
We shall denote this extension again by the same symbol. This extension maps $\ell^2(\Lambda_L^*)$ unitarily to $L^2(\Lambda_L)$
and is the inverse of $\hat{(\cdot)}|_{L^2(\Lambda_L)}$, see for example \cite[Theorem 8.20]{folland}.  From a physics background one might be familiar with the notation where  $L^2(\Lambda_L)$ represents the position space,  while $\ell^2(\Lambda_L^*)$ would be called momentum space,  or Fourier space.
For $ p \in \Lambda_L^*$  and $x \in \Lambda_L$ we define 
\begin{align} \label{eq:deofonb}
 \phi_p(x) ={|  \Lambda_L|}^{-1/2} e^{ i 2 \pi p \cdot x } .
\end{align}
Observe that 
$
\{ \phi_p :p \in \Lambda_L^* \} 
$
is an orthonormal Basis (ONB) of $L^2(\Lambda_L)$ \cite[Theorem 8.20]{folland} and that  by the definition  given in \eqref{FToff} 
$\hat{f}(p) = | \Lambda_L|^{1/2}  \inn{  \phi_p , f}_{L^2(\Lambda_L)} $ holds.

We are going to introduce the Laplacian with periodic boundary conditions on $L^2(\Lambda_L)$ by means of Fourier series.   
In this paper we shall adapt the physics convention that for a vector $a =(a_1,...,a_d)$ we use the notation $a^2 := |a|^2$ where $|a| :=  \sqrt{ \sum_{j=1}^d |a_j|^2}$.  
Using this notation,  we define the energy function
\begin{equation}
\label{energiyfunction}
\nu : \R^d  \to \R_+,  \quad p \mapsto \nu(p) :=  (2\pi p)^2,   
\end{equation} 
which we can now be used to
 define $-\Delta_L$ as the linear operator with domain 
\begin{align*}
D(-\Delta_L) = \{ f \in L^2(\Lambda_L) : \nu \hat{f} \in \ell^2(\Lambda_L^*) \},
\end{align*} 
and the mapping rule
\begin{equation} \label{eq:deoflaplac0} 
 -\Delta_L : D(-\Delta_L)  \to L^2(\Lambda_L), \quad f \mapsto -\Delta_L f   = ( \nu  \hat{f})^\vee . 
\end{equation} 
Observe that  $-\Delta_L$ is selfadjoint, since it is  unitary equivalent to  a multiplication operator by a real valued function,  and that we have the identity \
\begin{equation} \label{eq:deoflaplac} 
\left[ -  \Delta_L  f \right]^\wedge(p)  = \nu (p) \hat{f}(p)  .
\end{equation} 
%and that $-\Delta_L$ is essentially self-adjoint on the $L$-periodic smooth functions on $\R^d$ \cite[Theorem ??? ]{ReedSimon.1978}.
By  
\begin{equation} \label{defofmainH}  
 H_{L} := H_{\lambda,L} := - \frac{\hbar^2}{ 2 m } \Delta_L + \lambda V_{L}
\end{equation} 
we denote a random Schr\"odinger operator acting on $L^2(\Lambda_L)$ with   a random potential
$
V_{L} = V_{L,\omega}(x)
$, defined below,  and a  coupling constant $\lambda \geq 0$.   We choose units for the mass $m$ and Planck's constant  so that $\frac{\hbar^2}{2m} = 1$.
For a function $h : \R^d \to \C$ we denote by $h_{\#,L}$ or short $h_{\#}$ the $L$--periodic extension of $h|_{\Lambda_L}$ to $\R^d$,
explicitly for $y \in \R^d$ we define 
%For $\varphi , \psi  \in L^2(\R^d)$ we understand 
%\begin{align}
%\langle \varphi , \psi \rangle_L  := \langle \varphi|_{\Lambda_L} , \psi|_{\Lambda_L}  \rangle  .
%\end{align} 
\begin{align}
\label{defofperiodicext} 
h_{\#,L}(y) =   h|_{\Lambda_L}(x) ,
\end{align} 
where $x$ is the unique element in $\Lambda_L$ such that $x-y \in (L\Z)^d$,  i.e.,  $y = x + z$ for some $z \in  (L\Z)^d $. 
The potential is given by
\begin{align} \label{defofpot101} 
V_{L,\omega}(x) := \int_{\Lambda_L} B_{\#,L}(x - y) d \mu_{\omega}(y),
\end{align} 
where $B$, having the physical interpretation as   a single site potential profile and $\mu_{\omega}$ is a Poisson point process   on $\R^d$ with homogeneous unit density and 
with independent indentically distributed (i.i.d.) random masses.  
More precisely, for almost all realizations $\omega$, it consists of a countable,
locally finite collection of points $\{ y_\gamma(\omega) \in \R^d : \gamma =1,2,...\}$, and random weights
$\{ v_\gamma(\omega) \in \R : \gamma=1,2,...\}$ such that the random measure is given by
\begin{align} \label{EYS(2.3)}
\mu_\omega = \sum_{\gamma=1}^\infty v_\gamma(\omega) \delta_{y_\gamma(\omega)} , 
\end{align}  
where $\delta_a$ denotes the Dirac mass at $a \in \R^d$. The Poisson process $\{ y_\gamma(\omega) \}$ 
is independent of the weights $\{ v_\gamma(\omega)\}$. The weights are i.i.d. random variables with 
distribution ${\bf P}_v$ and with moments
\muun {\label{moments}
m_k :=  {\bf E}_v v_\gamma^k
}
 satisfying
%\begin{equation} \label{2.4} 
% m_{k } < \infty     , \quad  \text{ for all }  k \in \N.
%\end{equation} 

\begin{equation} \label{2.4d} 
 m_{k } < \infty     , \quad  \text{ for all }  k =1, 2,..., d+1. %\in \N.
\end{equation} 
%,   is  assumed  to be a  real valued Schwartz function on $\R^d$.
%Moreover, we assume that either $B$  has   compact support
%or that  $B$  is  symmetric with respect to the reflections of the coordinate axis,  i.e.  for  $j=1,...,d$
%\muun{ \label{SC}
%\mathfrak{S}_j:  (x_1,...,x_j,...,x_d) &\mapsto (x_1,...,-x_j,...,x_d) 
%\\ 
%B \circ \mathfrak{S}_j &= B. \nn
%}  
%\begin{remark}
%Each of the two conditions is mathematically convenient in the sense that they ensure sufficiently fast decay of the Fourier transform.  While the reflection symmetry condition is satisfied by rationally invariant potentials, which occur naturally.
%\end{remark}
We shall denote the expectation with respect to the above probability space by ${\bf E}$. 
In order to be able to control the growth of the potential, we assume that there exists a constant $C_B$ and an $\epsilon > 0$  such that 
\begin{align}  \label{propofprofileB} 
|B(x)| \leq C_B \langle x \rangle^{-d-1-\epsilon} ,  \quad x \in \R^d , 
\end{align} 
where we defined $\langle x \rangle := \sqrt{ 1 + |x|^2}$. 
Observe that we can write   \eqref{defofpot101} as 
% \tcr{Ommega-Abhängigkeit entweder auch links weglassen, oder auch rechts hinschreiben!}
\begin{equation}\label{bumpnotationres} 
 V_{L,\omega}(x) =  \sum_{\gamma : y_\gamma(\omega) \in \Lambda_L}  v_\gamma(\omega) B_{\#,L}(x-y_{\gamma}(\omega))  . % V_{L,\gamma}(x)   \quad \text{ with } \quad V_{L,\gamma}(x) := v_\gamma B_{\#,L}(x-y_{L,\gamma}) .
\end{equation} 

Furthermore one can see, that the potential \eqref{bumpnotationres}  is almost surely bounded and therefore it follows  by standard perturbation theorems, e.g. the Kato-Rellich theorem \cite{SR2},  that   the operator \eqref{defofmainH}  is almost 
surely self-adjoint  for all $\lambda \geq 0$.

We note that  the restriction of the random measure $\mu_\omega$  to the box $\Lambda_L$ has the same 
distribution as  a Poisson point process   on $\Lambda_L$ with homogeneous unit density and 
with independent indentically distributed  random masses with distribution ${\bf P}_v$. This fact can  be used 
to give  an equivalent more explicit definition of the random Hamiltonians $H_L$, as  outlined in
the following remark.

Let us now present the two main results concerning the Poisson distributed random potential, while the main technical work including the proofs of multiple smaller statements which will be necessary will be given in the succeeding chapters.
The first main result states the existence of the limit of the finite volume diagonal matrix elements $\langle \varphi ,  f( H_L ) \varphi \rangle_L  $ for $f \in C_\infty$ as $L \geq 1$ grows to infinity and reads as follows.

\begin{theorem} \label{feynmankac-3} For any $\varphi \in L^2(\R^d)$ and  $$f \in C_\infty(\R) := \{ f : \R \to \C : \text{continuous and vanishing at infinity} \}$$  the limit 
\begin{align} \label{poslinfunc-1} 
\lim_{L \to \infty} \langle \varphi ,  f( H_L ) \varphi \rangle_L  
\end{align} 
exists for almost all $\omega$. 
\end{theorem}

%\begin{proof}
%This follows from standard arguments from Theorem \eqref{resulforcompactsupport} and density of the functions $r_z : \R \to \C, x \mapsto (x-z)^{-1}$, for $z \in \C \setminus \R$,  in $C_\infty(\R) := \{ f : \R \to \C : \text{continous and vanishing at infinity} \}$. 
%\end{proof} 

%In the theorem below we will show   that it  follows  from the Portmanteau Theorem, cf.  Theorem \ref{smallPMT}, that the convergence is indeed weak, cf. Definition \ref{defofvagueweak}. 

%For the second main theorem of this work will consider the sequence of the respectiv spectral measures associated with the operators $H_L$ for $L>0$. Theorem \ref{feynmankac-3} as well as the Portmanteau Theorem grant the following.

The proof of Theorem \ref{feynmankac-3} will be given at the end of  Section \ref{proofofmainpoisson}. 
The theorem  allows us to define a limiting spectral measures  by means of the Riesz-Markov theorem.
This will be used  in  the following theorem. 

\begin{theorem} \label{ThmFolgPMTgen-0}  Let $\varphi \in L^2(\R^d)$. Then for almost all $\omega$ there exists a unique measure $\mu_{\varphi,\omega}$ on $\R$ such that  for all $f \in C_c(\R)$ 
\begin{align}
\lim_{L \to \infty} \langle \varphi ,  f(H_{L,\omega}) \varphi \rangle =  \int f(\lambda) d\mu_{\varphi,\omega}(\lambda) .
\end{align} 
The spectral measures of $H_{L,\omega}$ with respect to $\varphi |_{\Lambda_L}$, denoted by   $\mu_{\varphi,\omega,L}$, 
converge in the limit $L \to \infty$ vaguely and  weakly   to the measure $\mu_{\varphi,\omega}$, i.e.  
\begin{align}
\label{weakKonv}
\int_\R f(\lambda)  d \mu_{\varphi,\omega}(\lambda)  = \lim_{L \to \infty} \int_\R  f(\lambda) d   \mu_{\varphi,\omega,L}(\lambda).
\end{align}
for all  $f \in C_c(\R^d)$   (vague) and all bounded contiuous $f$ on $\R$ (weak),  respectively.
In particular,  all of the statements from the  Portmanteau Theorem \cite{Klenke.2014}  hold for the sequence $\mu_{\varphi,\omega,L}$.
\label{masskonv}
\end{theorem}

The proof of Theorem \ref{ThmFolgPMTgen-0}  will be given at the end of  Section \ref{proofofmainpoisson}.

\section{Linearly  bounded  potentials: resolvents} 
\label{proofs}
In this  section  we will show 
 a generalized version of  Theorem  \ref{feynmankac-3}, which holds for a larger class of  linearly bounded 
 potentials, i.e.,  the existence of the finite volume diagonal  matrix elements $\langle \varphi ,  f( -\Delta_L + W_L ) \varphi \rangle_L$ for $f \in C_\infty$ as $L \geq 1$ grows to infinity for a class of potentals $W_L$ satisfying a linear growth condition, which 
is stated in   Hypothesis \ref{hyp1}.  
The proof of Theorem \ref{feynmankac-3} is the main technical result of this work and requires some preparation. First, Proposition \ref{resulforcompactsupport} states a weaker version of Theorem \ref{mainThm} for compactly supported $\varphi$. To perform the proof of Proposition \ref{resulforcompactsupport}, we need a version of a resolvent identity, Lemma \ref{lem:abstractres}, as well as Lemma \ref{combesthomas}, which uses a Combes-Thomas argument (see \cite{CombesThomas.1973,  SR4}) to quantify the exponential decay of the resolvent.  

We start by introducing some natation. For $x \in \R^d$ we will use the notation %$| x | = \left( \sum_{i=1}^d x_i^2 \right)^\frac{1}{2}$  and 
$| x |_\infty := \max \{ x_i : 1 \leq i \leq d  \}$. % as well as $\langle x \rangle = \left( 1+ |x|^2 \right)^\frac{1}{2} $. \tcr{Die Klammerschreibweise wird schon in 4.9 benutzt}
Furthermore, 
we will use  the notation 
\begin{align} \label{notforops} 
p_L = - i \nabla  , \quad p_{L,s} = - i \partial_s ,  \quad s=1,...d,
\end{align} 
  with  domain $ \{ \psi \in H^1_{\rm loc}(\R^d) : \psi  \text{ $L$-periodic}\}$ corresponding to 
periodic boundary conditions for the operators on the right hand side. Furthermore, by 
the fact that the Fourier transform converts derivation to multiplication we can express $T_L$ in terms of derivatives 
\begin{align} \label{notforops-2} 
T_L = - \Delta_L  = \sum_{s=1}^d \partial_s^2  . 
\end{align} 
with domain 
$ \{ \psi \in H^2_{\rm loc}(\R^d) : \psi  \text{ $L$-periodic}\}$. 

We shall base the result on the following working Hypothesis, for which we will show that it is satisfied for the random potential in Lemma \ref{vldifflarge} later.

\begin{hyp} \label{hyp1}  The sequence  $W_L$ of $L$-periodic real valued  measurable  functions on $\R^d$.
And satisfies the following. 
\begin{itemize}
\item[(i)]  There exists a constant $C$ such that  
\begin{align}
 | W_L(x ) | \leq C \langle x \rangle .
\end{align}
\item[(ii)]  
 Let $ O  = \Lambda_{1/2}$. Then there exits a   constant $C$ and positive $\alpha, \epsilon > 0 $ such that for all sufficiently large  $ 1\leq  L \leq L'$ 
\begin{align}
\sup_{x \in \Lambda_{L}} |  1_{L^{1/2+\alpha} O}(x)   ( W_L(x) - W_{L'}(x)  ) |  \leq C  L^{-\epsilon}  .
\end{align}
\end{itemize} 
\end{hyp} 

The main technical result  to prove Theorem  \ref{feynmankac-3}   is based on the  following theorem.  The proof of Theorem \ref{mainThm} will be given near the end of this section, since it requires some preparation first.

\begin{theorem} \label{mainThm} Suppose Hypothesis \ref{hyp1} holds.  
 Let $\varphi \in L^2(\R^d)$. 
Then for any $z \in \C \setminus \R$ the limit
\begin{align} \label{maintech1}
\lim_{L \to \infty} \langle  \varphi , ( T_L + W_L - z )^{-1} \varphi \rangle_L 
\end{align} 
exists.  In fact, for any $f \in C_\infty(\R)$ the following  limit exists  
\begin{align}\label{maintech2}
\lim_{L \to \infty} \langle  \varphi , f  ( T_L + W_L)  \varphi \rangle_L  .
\end{align} 
\end{theorem}

To work towards the proof of  Theorem \ref{mainThm} we consider the following proposition, which is a weaker version of Theorem \ref{mainThm}  for compactly supported $\varphi$. 

\begin{proposition} \label{resulforcompactsupport} 
  Suppose Hypothesis \ref{hyp1} holds.  
Let $\varphi \in L^2(\R^d)$, which vanishes outside of a ball of radius $R > 0$.
Then for any $z \in \C \setminus \R$ the limit
\begin{align}
\lim_{L \to \infty} \langle  \varphi , ( T_L + W_L - z )^{-1} \varphi \rangle_L 
\end{align} 
exists. 
\end{proposition}

For the proof of Proposition  \ref{resulforcompactsupport} we will need two different Lemmas \ref{lem:abstractres} and  \ref{combesthomas}  starting with the
following one, giving a variant of a resolvent identity. 

\begin{lemma} \label{lem:abstractres}   Let $A$ and $B$  be closed  densely defined operators  in Hilbert spaces $\hh_1$ and $\hh_2$, 
respectively, which have bounded inverse. 
 Let $J : \hh_1 \to \hh_2$ be a linear map which maps  the  domain of $A$ into the domain of $B$. 
Then 
\begin{align} \label{secondresident} 
B^{-1} (  B J - J  A )  A^{-1} = J  A^{-1}   - B^{-1} J .
\end{align} 
\end{lemma} 
\begin{proof}
This statement follows from a simple calculation.
\end{proof}

%\tcr{überleitung zum Lemma? As we want to show that  $\langle  \varphi , ( T_L - W_L - z )^{-1} \varphi \rangle_L $   for the restriction to $\Lambda_L$ is a Cauchy sequence for  $L \to \infty$, we need to be able to compair different box sizes.  The following Lemma \ref{vldifflarge} therefor calculates the difference of the effect of the potential between two different box lengts $L \leq L'$. For some intuition 
% the potentials  differ in the poisson scatter points that are located in the area $\Lambda_{L' } \setminus \Lambda_{L} $. Therefore for large $L>0$ the difference of the effect of the potentials is small for points near the origine since  $\Lambda_{L' } \setminus \Lambda_{L} $ is far and the bumbs have already seen much decay.  So a differenz will be large  far from the origine, which we can compensate with the resolvent being small in this area.  Idee von $ O  = \Lambda_{1/2}$ und $1_{L^{1/2+\alpha} O}$ kurz erklären}
%
%
%\tcr{überleitung! The following Lemma \ref{combesthomas} will now apply the previous Lemma \ref{vldifflarge} to.....}

The following lemma quantifies the exponential decay of the resolvent.
\begin{lemma} \label{combesthomas}
 Suppose Part (i) of Hypothesis \ref{hyp1} holds. 
%  Suppose $|V(x) | \leq C_0 \langle x \rangle $ for some constant $C_0$.
Suppose that $\varphi$ 
has compact support in $\Lambda_{L_0}$ for some $L_0 \geq 1$.  Let $O$ be an open neighborhood of zero such that $\overline{O} \subset (-1/2,1/2)^d$. Let $\alpha \in (0,1/2)$. Let $z \in \C\setminus \R$.  Then there exist positive constants $\kappa$ and $C_0$ 
such that  for $L \geq L_0$
\begin{align*} 
&  \| 1_{L^{1/2 + \alpha } O^c}    (T_{L} + W_{L} -  z  )^{-1}   \varphi \|  \leq C_0 \exp( - \kappa L^{\alpha } )  \sup_{x  \in \supp \varphi} e^{   \langle x  \rangle  } \|   \varphi  \|.
\end{align*}  
\end{lemma} 

\begin{proof}
To show spacial decay of the resolvent we   use a method known as  Combes-Thomas argument, see  \cite{SR4}
and references therein.
% \tcr{sollte man hier eine referenz angeben, Reed Simon Band 4?}
%We use $\eta(x) = \int \varphi(a)  \langle x  - a \rangle_{\#,L} da $,  where 
%$\varphi \in C_c^\infty(\R^d)$   with support in the unit ball and $\int_{\R^d} \varphi(y) dy = 1$.
%Observe that $\nabla \langle x \rangle = \frac{x}{|x|}$ and so   
We define  
\begin{align} 
\label{defeta} \tilde{\eta}_L(x) =L \psi \left( \frac{\langle x \rangle}{L} \right) ,
\end{align} 
 where we assume that $\psi : [0,\infty)  \to \R$ is smooth and has the 
following properties 
\begin{align}
& \psi(r)  = r \text{ ,  if } r \in \left[0,\frac{1}{4}\right]  , \label{const1}  \\
& 0 \leq \psi'(r) \leq 1  \text{ ,  if } r \in \left[\frac{1}{4},\frac{3}{8}\right]   ,  \label{const1-0}  \\
& \psi(r) = \const   \text{ ,  if } r \in \left[\frac{3}{8},\infty\right] \label{const3}.
\end{align} 
It is basic to see that this function $\psi$  satisfies  
\begin{align} 
\label{boundonpsi} \psi(r)  \leq r . 
\end{align} 
For example such a function can be obtained via $
\psi(r) =  \int_0^r  \left( \varphi  * 1_{\left[-\frac{5}{16},\frac{5}{16}\right]} \right)(s)  ds,
$
where $0 \leq \varphi $ is a smooth function with support in $[-\frac{1}{100}, \frac{1}{100}] $ and $\int \varphi(t)d t = 1$.
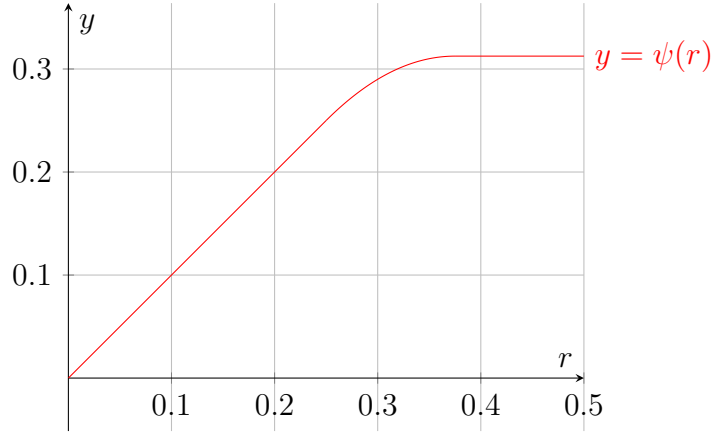
\begin{center}
  \begin{tikzpicture}
    \begin{axis}[axis lines = left, axis equal, grid=both, axis x line=middle, axis y line=middle, xlabel=$r$, ylabel=$y$, clip=false]
    %  \addplot[domain=0:2,color=blue, samples=50] {x^2} node[right] {$f(x) = x^2$};
      \addplot[domain=0:0.25,color=red, samples=100] {x} ;
            \addplot[domain=0.25:0.375,color=red, samples=100] {0.25+0.0625-4*(x-0.375)^2)} ;
          \addplot[domain=0.375:0.5,color=red, samples=2]{0.25+0.0625} node[right] {$ y = \psi(r)$} ;
    \end{axis}
  \end{tikzpicture} 
\end{center}
% \tcr{hier würde ein Bild von $\varphi$ helfen. Evtl einfach die concrete Funktion angeben?} The function  $\psi$ is  linear \tcr{decreasing?} from $0$ to $1/4$ with slope one 
%and from $1/4$
%to $3/8$ it increases with slope less or equal to one until it is constant after  $3/8$.  
We will study the $L$-periodic extension of $\tilde{\eta}_L |_{\Lambda_L}$ cf. \eqref{defofperiodicext} 
\begin{align} \label{etadef} 
\eta_L(x) = \tilde{\eta}_{L,\#}(x) . 
\end{align} 
%For notational convenience we shall occasionally write $\eta$ for $\eta_L$.  \tcr{ist das notwendig?}
We will show that for all $x \in \Lambda_L$ we have

 \begin{align}  
 & c_d \langle x \rangle \leq \eta_{L}(x)  \leq \langle x \rangle    , \qquad c_d := 1 / ( 8  \sqrt{d}).\label{boundofdecayfunc}   
\end{align} 
First we observe that the second inequality follows immediately from \eqref{boundonpsi}
 and hence $\tilde{\eta}_L(x) \leq \langle x \rangle$. 
To show the first inequality in  \eqref{boundofdecayfunc}   we will use  the elementary inequalitities 
\begin{align} 
& | x |_\infty \leq |x| \leq \sqrt{d} | x |_\infty  \label{geom1}  \\
&
| x | \leq \langle x \rangle \leq 1 + | x |.  \label{geom2} 
\end{align} 
By the assumed properties  \eqref{const1}--\eqref{const3}  of  $\psi$, it follows  $\psi(s) \geq   s / ( 8  \sqrt{d})$ whenever  $0 \leq  s \leq 2 \sqrt{d}$. 
Thus it follows that $\tilde{\eta}_L(x) = L \psi( \langle x \rangle / L) \geq \langle x \rangle  / ( 8  \sqrt{d})$
whenever $\langle x \rangle \leq L 2 \sqrt{d}$. 
Since implies  
For  $x \in \Lambda_L$ by  \eqref{geom1} and \eqref{geom2}  for $L \geq 1$ we have
\begin{align*}
\langle x \rangle \leq 1 + |x|
\leq 1+\sqrt{d}|x|_{\infty} \leq  1+ \sqrt{d} L/2\leq  L 2 \sqrt{d}.
\end{align*} Now it follows that 
 \begin{align} 
 \frac{ \langle x \rangle  }{  8  \sqrt{d} }  \leq \tilde{\eta}_L(x)  
\end{align} 
for all $x \in \Lambda_L$. This shows the first inequality in  \eqref{boundofdecayfunc}.  

% \begin{align} 
% & c_d \langle x \rangle \leq \eta_{L,\#}(x)  \leq \langle x \rangle    , \qquad c_d := 1 / ( 8  \sqrt{d})  \label{boundofdecayfunc}   
%\end{align} 
Next we estimate the derivative of $\eta_L$. 
By  construction of $\psi$ we have  $0 \leq \psi'\leq 1$  and so 
\begin{align} \label{estonetatilde}  | \nabla \tilde{\eta}_L(x) |  = \left| \psi' \left( \frac{\langle x \rangle}{L}  \right)  \frac{x}{ \langle x \rangle} \right| \leq 1. \end{align}

It follows from the construction that  $\eta_L$ is smooth, since  in a neighborhood of the boundary of $\Lambda_L$,  it follows that $\frac{3}{8} L \leq | x |_\infty \leq \langle x \rangle$
 (by \eqref{geom1}) and so $\tilde{\eta}_L(x)$ is constant in view of \eqref{etadef},  \eqref{defeta},  and \eqref{const3}.  Thus, we conclude  from \eqref{estonetatilde} 
 for $x \in \R^d$ 
% \tcr{Gleichungen angeben aus welchen das folgt}
 \begin{align} 
 & | \nabla \eta_L(x) |  \leq 1 . \label{gradboundofdecayfunc} 
\end{align}

%We use $\eta(x) = \int \varphi(a)  \langle x  - a \rangle_{\#,L} da $,  where 
%$\varphi \in C_c^\infty(\R^d)$   with support in the unit ball and $\int_{\R^d} \varphi(y) dy = 1$.
%Observe that $\nabla \langle x \rangle = \frac{x}{|x|}$ and so   
Next we introduce an analytic extension of the Hamiltionian.  For this, we will write $e^{- i \alpha \eta_L} $, referring to the multiplication operator associated to this function.
Recall  \eqref{notforops}.
Then  using that  for $\alpha \in \R$ and   $\phi \in D(p_L)$  we have 
$$p_L e^{- i \alpha \eta_L} \phi = - i \nabla (  e^{- i \alpha \eta_L} \phi  ) 
= - ( \alpha \nabla \eta_L ) (  e^{- i \alpha \eta_L} \phi  )  + e^{- i \alpha \eta_L}  p_L \phi  
=e^{- i \alpha \eta_L}  ( - \alpha \nabla \eta_L + p_L) \phi $$
we find with $K_L := T_L + W_L$ that 
\begin{align}\label{anafamily} \nn
 e^{ i \alpha \eta_L}  K_L e^{ - i \alpha \eta_L}  
 &=  e^{ i \alpha \eta_L}  ( p_L^2 + W_L )  e^{ - i \alpha \eta_L}
 =  ( e^{ i \alpha \eta_L}  p_L   e^{ - i \alpha \eta_L})^2 + W_L \nn \\
 &= 
  ( p_L - \alpha \nabla \eta_L)^2 
 + W_L 
 = T_L  -     \alpha  ( p_L \cdot  \nabla \eta_L +  \nabla \eta_L \cdot p_L )   + \alpha^2 ( \nabla \eta_L) ^2  + W_L 
 \nn \\  
 & =: K_L(\alpha)  .
\end{align} 
Now the right hand side of  \eqref{anafamily} also makes sense for  $\alpha \in \C$. 
In fact as a function of $\alpha$  \eqref{anafamily}  is an  analytic family of type (A)  on $\C$ \cite{SR4},
since $p_L$ is infinitesimally small with respect to $-\Delta_L$, cf.  \cite{SR2}  and  $\eta_L$ is smooth with bounded derivatives. 
It follows immediatly from  the definition  that for $\alpha ,\beta \in \R$ 
%\tcr{das gefällt mir nciht,  $\alpha$ ist manchmal komplex und manchmal reell.  Ergibt  $\alpha + \beta $ einen Sinn wenn beides reell ist?}
\begin{align} \label{unitaryop} 
K_L(\alpha + \beta )  =   e^{ i  \beta  \eta_L} K_L(\alpha )  e^{- i \beta \eta_L} 
\end{align} 
and so by analytic continuation \eqref{unitaryop} holds also for $\alpha = i a$ with $a \in \R$. 
Thus  $K_L(i a + \beta )$ and  $K_L(i a )$ are unitarily equivalent. 
 In particular for  % $\alpha = i a $ with 
$a \in \R$, we find  
\begin{align}
\label{HLia}
K_L(i a ) = - \Delta_L -  i a  ( p_L \cdot \nabla \eta_L + \nabla \eta_L  \cdot p_L ) - a^2 ( \nabla \eta_L)^2 + W_L . 
\end{align}

We want to obtain bounds on the resolvent of this operator. 
To this end we  consider its  numerical range. Let  $\varphi \in D(-\Delta_L)$. Then  by positivity of the Laplacian,  (i.e.  $\langle \varphi , - \Delta_L \varphi \rangle \geq 0$) and \eqref{gradboundofdecayfunc} we find for the real expectation value of  \eqref{HLia} with respect to $\varphi$  that
for $a \in \R$  
\begin{align} \label{eq:estonrange} 
{\rm Re} \langle \varphi , K_L( i a ) \varphi \rangle \geq - a^2 + \inf W_L  .
\end{align} 
From   \eqref{gradboundofdecayfunc}, giving us $ a^2 ( \nabla \eta_L)^2 \leq a^2$, 
 %and using the positivity of the Laplacian again, 
 we find using \eqref{HLia}
%\tcr{den dritten Schritt ausfuhrlicher?}
\begin{align*}
& |{\rm Im} \langle \varphi , K_L( i a ) \varphi \rangle | 
\\
%&=| \langle \varphi , K_L( i a ) \varphi   \rangle - {\rm Re} \langle \varphi , K_L( i a ) \varphi \rangle  |
%\\ &\leq |\langle \varphi , ( [- \Delta_L -  i  a ( p_L \cdot  \nabla \eta_L + \nabla \eta_L  \cdot p_L ) - a^2 ( \nabla \eta_L)^2 + W_L] - [- a^2 + \inf W_L] )  \varphi \rangle  |
%\\ 
&  = |a (  \langle p_L \varphi , \nabla \eta_L \varphi \rangle  + \langle \nabla \eta_L \varphi , p_L \varphi \rangle  )  |  %\tcr{\text{ist es wichtig, dass $p$ auf der anderen Seite steht?}}
\\
& \leq |a| 2 \| |p_L|  \varphi \| \| \varphi \|  \\
& = |a| 2 \sqrt{ \langle \varphi ,  p_L^2 \varphi \rangle } \| \varphi \|   
= |a| 2 \sqrt{ {\rm Re} \langle \varphi , ( K_L(i a)+  a^2  |\nabla \eta_L|^2 - W_L) \varphi \rangle   } \| \varphi \|\\
& \leq |a| 2\sqrt{ {\rm Re} \langle \varphi , ( K_L(i a)+  a^2  - \inf W_L) \varphi \rangle   } \| \varphi \| . 
\end{align*} 

\marginpar{Ende 2024-03-21} 
Thus we find for $x  + i y = \langle \varphi , K_L( i a ) \varphi \rangle $  with $\| \varphi \| =1$ that 
\begin{align} \label{estony} 
|y| \leq |a| 2 \sqrt{ x  + a^2  - \inf W_L } . 
\end{align} 
This characterizes the numerical range. 
Thus by  \eqref{eq:estonrange}  and \eqref{estony}  we can estimate the numerical range  as follows  
$$
{\rm nr} K_L(ia) \subset \{ x  +  i y :  x \geq \inf W_L - a^2  , |y| \leq |a| 2 \sqrt{ x  + a^2  - \inf W_L } \}  =: \mathcal{Q}_{L,a} . 
$$
\begin{center} 
\begin{tikzpicture}[scale=0.8]
  \draw[->] (-5, 0) -- (5, 0) node[right] {$\R$};
  \draw[->] (0,  - 4) -- (0,3) node[above] {$i \R$}; 
%\filldraw[blue]  (1,2)  circle[radius=2pt] node[right] {$z$};
\draw[-, thick] ( -3 , 0.1 ) -- ( -3 , -0.1 )  node[below] {$-a^2 + \inf W_L $} ;
\fill  (-0.3,0);
          \filldraw[ domain=-2:2, smooth, variable=\y, red, opacity =0.2 ]  plot ({2*\y*\y-3}, {0.5*\y});
          \draw[ domain=-2:2, smooth, variable=\y, red]  plot ({2*\y*\y-3}, {0.5*\y}) node[above]{$\mathcal{Q}_{L,a}$};
\end{tikzpicture}
\end{center}

Now by  Hypothesis   \ref{hyp1}  we have $W_L(x) \geq - C_V L$ for some $C_V \geq 1$ and large $L \geq 1$. 
To estimate the distance from the numerical range  we first partition the set $\mathcal{Q}_{L,a}$, then we use  the triangle inequality, and finally  use  
the definition of the set   $\mathcal{Q}_{L,a}$
\begin{align*}
 & {\rm dist}(z  , {\rm nr} K_L(ia) ) \\
&  \geq \inf_{x+i y \in \mathcal{Q}_{L,a}} \left| z  - x - i y \right| \\
& = \min\left\{  \inf_{x+i y \in \mathcal{Q}_{L,a}: \left|x- {\rm Re} z \right| \geq 1} \left| z  - x - i y \right| ,   \inf_{x+i y \in \mathcal{Q}_{L,a}: \left|x- {\rm Re} z \right| \leq 1} \left| z  - x - i y \right| \right\} \\
& \geq \min\left\{ 1 ,   \inf_{x+i y \in \mathcal{Q}_{L,a}: \left|x-{\rm Re} z \right| \leq 1} \left|{\rm Im} z \right|  -  \left| y \right| \right\} \\
& \geq \min\left\{ 1 ,  \left|{\rm Im} z \right| -  \sup_{x+i y \in \mathcal{Q}_{L,a}: \left|x-{\rm Re} z \right| \leq 1}      |a| 2 \sqrt{ x  + a^2  - \inf W_L } \} \right\} \\
& \geq \min\left\{  1 ,     \left| {\rm Im} z \right| - \left|a\right| 2 \sqrt{ {\rm Re} z  + 1 + a^2 +  C_V L   } \right\}  \\
& \geq \min\left\{  1 ,     \left| {\rm Im} z \right| /2  \right\} , 
\end{align*}
where the last inequality holds provided 
 \begin{align}
  \label{choicofa} |a | \leq    \min\left\{ 1 , \frac{  | {\rm Im} z |}{4  \sqrt{   {\rm Re} z  + 2 + C_V L}} \right\} . 
 \end{align}  
Thus for $a \in \R$ satisfying   \eqref{choicofa} and $b \in \R$  we find using  \eqref{unitaryop} 
\begin{align} \label{boundonres} 
\| (K_L(ia + b ) - z  )^{-1}  \| \leq  \frac{1}{{\rm dist}( z  , {\rm nr}   (K_L(ia) ) }  \leq  \max \left\{ 1 , \frac{2}{|{\rm Im} z|}  \right\} 
\end{align}

%\tcr{Until here on 2024-06-11???} Thus going back to  \eqref{eq:secondres}  we find for $L \geq L'$   and  $a \geq 0$ satisfying    \eqref{choicofa}
Now we are ready  to estimate the norm of the resolvent applied to $\varphi$  and multiplied with some    function $\tilde{\chi}$  with support in the interior of $\Lambda_L$
\begin{align} 
\| \tilde{\chi}  (T_{L} + W_{L} -  z  )^{-1} \varphi \| .
\end{align} 
For this we first observe that for real $\alpha$  %and $z \in \C$ with ${\rm Im} z \neq 0$
we have by unitarity and  the definition introduced in \eqref{anafamily} 
\begin{align}  \label{anaext1} 
&  \tilde{\chi }   (T_{L} + W_{L} -  z  )^{-1}   \varphi \nn \\
& =  \tilde{\chi } e^{- i \alpha  \eta_L} e^{i \alpha \eta_L}  (T_{L} + W_{L} -  z  )^{-1}  e^{- i \alpha \eta_L} e^{i \alpha \eta_L}   \varphi \nn  \\
& =  \tilde{ \chi } e^{- i \alpha  \eta_L} (  K_L(\alpha)  -  z  )^{-1}   e^{i \alpha \eta_L}   \varphi  .   
\end{align} 
Now the right hand side as a function of $\alpha$ can be analytically continued into a strip
around the real axis  as long as $a=- {\rm Im}\alpha$ satisfies   \eqref{choicofa}, and in which case we find  with $\alpha =-i a$ 
\begin{align} \label{ctsecondthree}
&  \tilde{\chi}    (T_{L} + W_{L} -  z  )^{-1}   \varphi  
%& =   \chi e^{- i \alpha  \eta} e^{i \alpha \eta}  (-\Delta_{L'} - W_{L'} -  z  )^{-1}  e^{- i \alpha \eta} e^{i \alpha \eta}   \varphi  \\
%& =   \chi e^{- i \alpha  \eta} (  K_L(\alpha)  -  z  )^{-1}   e^{i \alpha \eta}   \varphi  \\
 =
 \tilde{\chi}  e^{- a \eta_L}  (K_L(-i a)  -  z  )^{-1}    e^{ a \eta_L}   \varphi  
\end{align} 
Now calculating the norm of \eqref{ctsecondthree} and using   \eqref{boundonres}  as well as  the elementary bounds \eqref{boundofdecayfunc}  
\begin{align} 
&  \| \tilde{\chi}    (T_{L} + W_{L} -  z  )^{-1}   \varphi \|  \nonumber\\
& \leq 
 \| \tilde{\chi}  e^{- a \eta_L} \| \|   (K_L(-i a)  -  z  )^{-1}  \| \|   e^{ a \eta_L}   \varphi  \| \nonumber\\
& \leq  \| \tilde{\chi}  e^{-  a c_d \langle \cdot  \rangle} \|  \max\left\{ 1 , \frac{2}{|{\rm Im} z|}  \right\} 
 \sup_{x \in \supp \varphi }   e^{ a  \langle x  \rangle  } \|   \varphi  \| . \label{bounonresct} 
\end{align} 
Now we let 
$$
a= \frac{| {\rm Im} z |}{4  \sqrt{   {\rm Re} z  + 2 + C_V L}}.
$$
Then for $L$  sufficiently large we  find 
\begin{align}\label{choiceofa}
 \frac{| {\rm Im} z |}{8 \sqrt{    C_V L}} \leq  a \leq1
\end{align}
and hence \eqref{choicofa} holds. Next we choose  $\tilde{\chi} =   1_{L^{1/2 + \alpha } O^c} $ 
and let  $d_O := \inf\{ |x| : x \in O^c \}$. By assumption on the set $O$ we have $ 1_{L^{1/2 + \alpha } O^c}$ has support in $\Lambda_L$ and  $d_O>0$. Inserting this into \eqref{bounonresct} and using \eqref{choiceofa} as well as  \eqref{geom2} 
we find for large $L$ satisfying \eqref{choiceofa}
%Since for large $L$ we have $a \geq \frac{|{\rm Im} z|}{8 \sqrt{ C L}}$ by   \eqref{choicofa}.
%Thus as $L \to \infty$ we find assuming that $\tilde{\chi} =   1_{L^{1/2 + \alpha } O^c} $  we obtain
%with $d_O := \inf\{ |x| : x \in O^c \} > 0$  recalling \eqref{boundofdecayfunc}  
\begin{align*} 
&  \| \tilde{\chi}    (T_{L} + W_{L} -  z  )^{-1}   \varphi \|  \\
& \leq  \exp \left(  - c_d \frac{|{\rm Im} z |}{8 \sqrt{ C_V L}} d_{O}  L^{1/2 + \alpha}  \right) 
\max\left\{ 1 , \frac{2}{|{\rm Im} z|}  \right\} 
 \sup_{x \in \supp \varphi }   e^{   \langle x  \rangle  } \|   \varphi  \| .
\end{align*} 
Now this estimate implies the claim.
%Now this tends to zero as $L \to \infty$. In fact it decays of the order $\exp( - c' \sqrt{L})$
%for some $c' >$ and therefore controls polynomialy growing 
%expressions.\eqref{diffest1} \eqref{diffest1m} 
\end{proof} 

Now, with the resolvent identity from Lemma \ref{lem:abstractres} and Lemma \ref{combesthomas}, giving a decay estimate,  we are equiped to    prove  Proposition   \ref{resulforcompactsupport}.

\begin{proof}[Proof of Proposition   \ref{resulforcompactsupport}] We want to show that $\langle  \varphi , ( T_L + W_L - z )^{-1} \varphi \rangle_L$
is a Cauchy sequence. Assume that $L' \geq L$ and $\Lambda_L \supset B_R := \{ x \in \R^d : |x| \leq R\}$. 
Choose $\chi \in C_c^\infty$ with 
\begin{align}
\label{chichi}
\text{$0 \leq \chi \leq 1$ and 
$\chi(x)=1$, if $|x| \leq 1/4$, $\chi(x) = 0$, if $ |x| \geq 1/2$}. 
\end{align}
Define 
\begin{align}
\label{chiL}
\chi_L(x) = \chi(x/L). 
\end{align}
Now we want to apply Lemma  \ref{lem:abstractres}  for  the resolvents acting in the Hilbert spaces 
$L^2(\Lambda_L)$ and $L^2(\Lambda_{L'})$ and for $J  :L^2(\Lambda_{L}) \to L^2(\Lambda_{L'}) $ given by  multiplication with 
the  function $\chi_L$, i.e., 
\begin{align*}
(J \varphi)(x) := \begin{cases}
\chi_L(x)\varphi(x)  & \text{for} \qquad x \in   \Lambda_{L}
\\
0  & \text{for} \qquad x \in \Lambda_{L'} \setminus \Lambda_{L}.
 \end{cases}
\end{align*}
Using Lemma \ref{lem:abstractres}
 and  calculating a commutator we obtain % by means of the product rule of differentiation we find 
\begin{align}
 & \chi_L ( T_L + W_L  - z )^{-1} -  ( T_{L'} + W_{L'} - z  )^{-1}  \chi_L \nn \\
 &  =  ( T_{L'} + W_{L'} - z )^{-1} \left(  ( T_{L'} + W_{L'}-z) \chi_L - \chi_L  ( T_{L} + W_{L} -z)  \right) 
 ( T_{L} + W_{L} - z  )^{-1} \nn \\
  &  =  ( T_{L'} + W_{L'} -  z  )^{-1} \left(  -2 ( \nabla  \chi_L ) \cdot  i p_L  -   ( \Delta \chi_L  ) +  (W_{L'} - W_{L} )  \chi_L  \right) 
 ( T_{L} + W_{L} - z )^{-1} , \label{resChi}
\end{align} 
where we  calculated the commutator by means of the product rule, i.e.,  for $\psi$ in the  domain we have 
$$
T_{L'} \chi_L \psi = - ( \Delta \chi_L ) \psi - 2   ( \nabla \chi_L ) \cdot i p_L \psi + \chi_L  T_L  \psi ,
$$
recalling  \eqref{notforops} and   \eqref{notforops-2}. 
Calculating the inner product with $\varphi$ we obtain for $L \geq 4 R$ using \eqref{chichi}, in particular that   $\chi_L$ is one where
 $\varphi$  does not vanish, and \eqref{resChi} 
\begin{align}
 & | \langle \varphi ,    ( T_L + W_L  - z )^{-1} \varphi \rangle_L  - \langle \varphi ,   ( T_{L'} - W_{L'} - z  )^{-1} \varphi \rangle_{L'}  |  \nn \\
 & =| \langle \varphi ,  \left[ \chi_L ( T_L + W_L  - z )^{-1} -  ( T_{L'} - W_{L'} - z  )^{-1}  \chi_L 
\right] \varphi \rangle_{L'}  | \nn  \\
  &  = |  \langle \varphi ,  ( T_{L'} - W_{L'} -  z  )^{-1} \left( - 2 ( \nabla  \chi_L ) \cdot  i p_L -   ( \Delta \chi_L  ) +  (W_{L'} - W_L ) \chi_L\right) 
 ( T_{L} - W_{L} - z )^{-1}  \varphi \rangle_{L'}  | \nn \\
 &\leq |  \langle \varphi ,  ( T_{L'} - W_{L'} -  z  )^{-1}  ( \Delta \chi_L  )  
 ( T_{L} - W_{L} - z )^{-1}  \varphi \rangle_{L'}  | \nn \\
 & 
 + \sum_{s=1}^d \|   2 ( \nabla  \chi_L )_s   ( T_{L'} - W_{L'} -  z  )^{-1} \varphi \|_{L'} 
 \|  i (p_L)_s   ( T_{L} - W_{L} - z )^{-1}  \varphi \|_{L}\nn \\
 & + |  \langle \varphi ,  ( T_{L'} - W_{L'} -  z  )^{-1}  (W_{L'} - W_L )  \chi_L 
 ( T_{L} - W_{L} - z )^{-1}  \varphi \rangle_{L'}  | 
 ,\label{diffest1} 
\end{align} 
where in the last inequality we used the triangle inequality and Cauchy-Schwarz. 
Now the first term on the right hand side of \eqref{diffest1} is estimated using   \eqref{chiL} and the chain rule to conclude $\| \Delta \chi_L \|_\infty
\leq \| \chi'' \|_\infty L^{-2}  $, as well as \eqref{ln-1}
\begin{align}
& |  \langle \varphi ,  ( T_{L'} - W_{L'} -  z  )^{-1}  ( \Delta \chi_L  )  
 ( T_{L} - W_{L} - z )^{-1}  \varphi \rangle_{L'}  |  \nn \\
&  \leq \| \chi'' \|_\infty L^{-2} \frac{1}{|{\rm Im} z|^2} \| \varphi \|^2  \to 0 
\end{align} 
as $L \to \infty$. 

On the other hand,  the  bound of the  second term on the RHS of \eqref{diffest1} is obtained  giving estimates on the two occurring norms.  First we estimate 
\begin{align}
 \|    2 ( \nabla  \chi_L )_s   ( T_{L'} - W_{L'} -  z  )^{-1} \varphi \|_{L'} \leq 
2  \| \chi' \|_\infty L^{-1}  \frac{1}{|{\rm Im} z|} \|  \varphi \|    \label{diffest1mm} ,
\end{align}  
 which follows again from the spectal theorem, \eqref{ln-1} and $\| \nabla_s \chi_L \|_\infty
\leq \| \chi' \|_\infty L^{-1}  $ again using   \eqref{chiL} and the chain rule.  
The other norm can be estimated by using  \eqref{notforops-2} and the triangle inequality, as well as Hypothesis \ref{hyp1} (i) as follows
\begin{align}
&  \sum_{s=1}^d \|    p_{L,s}  (-T_{L} - W_{L} -  z  )^{-1} \varphi \|^2 \nn \\
&=\sum_{s=1}^d \langle  p_{L,s}  (-T_{L} - W_{L} -  z  )^{-1} \varphi ,    p_{L,s}  (-T_{L} - W_{L} -  z  )^{-1} \varphi \rangle \nn \\
&= \left\langle    (-T_{L} - W_{L} -  z  )^{-1} \varphi ,  - \sum_{s=1}^d \partial_s^2 (-T_{L} - W_{L} -  z  )^{-1} \varphi \right\rangle \nn \\
& =   \langle (T_{L} - W_{L} -  z  )^{-1}\varphi  ,   T_L    (T_{L} - W_{L} -  z  )^{-1} \varphi \rangle    \nn \\ 
& \leq  |   \langle (T_{L} - W_{L} -  z  )^{-1}\varphi  ,  ( T_L + W_L - z )   (T_{L} - W_{L} -  z  )^{-1} \varphi \rangle  |  \nn \\
&  \quad +  |   \langle (T_{L} - W_{L} -  z  )^{-1}\varphi  ,   (  W_L +  z )   (T_{L} - W_{L} -  z  )^{-1} \varphi \rangle  | \nn \\
& \leq \| \varphi \|^2  ( {\rm Im} z)^{-1} +  ( |z| + C \langle L \rangle )  \| \varphi \|^2  ( {\rm Im} z)^{-2}. \label{diffest1m} 
\end{align}  
Thus it follows from \eqref{diffest1mm} and  \eqref{diffest1m} that 
\begin{align}
 &  \|   2 ( \nabla  \chi_L )_s   ( T_{L'} - W_{L'} -  z  )^{-1} \varphi \|_{L'} 
 \|  i p_{L,s}   ( T_{L} - W_{L} - z )^{-1}  \varphi \|_{L} \nn  \\
& \leq 2  \| \chi' \|_\infty L^{-1}  \frac{1}{|{\rm Im} z|^2}  \| \varphi \|^2   \left( ( {\rm Im} z) +  ( |z| + C \langle L \rangle )   \right)^{1/2} 
\to 0 
\end{align} 
as $L \to \infty$.  
Let us now consider the last term on the RHS of \eqref{diffest1}. For $\alpha \in (0,1/2)$
we  estimate 
%\tcr{We will now use \eqref{chiL} in particulare its compact suport to 
%be able to evaluate the $L'$-Norm of the $L-Operator$ times $ \chi_L 
% ( T_{L} - W_{L} - z )^{-1}  \varphi$.}
\begin{align}
 & |  \langle \varphi ,  ( T_{L'} - W_{L'} -  z  )^{-1}   (W_{L'} - W_L )  \chi_L
 ( T_{L} - W_{L} - z )^{-1}  \varphi \rangle_{L'}  | \nonumber \\
& \leq |  \langle \varphi ,  ( T_{L'} - W_{L'} -  z  )^{-1}  1_{L^{1/2+\alpha}  \Lambda_{1/2} }   (W_{L'} - W_L )   \chi_L
 ( T_{L} - W_{L} - z )^{-1}  \varphi \rangle_{L'}  | \label{terminrhsest1}  \\
& + |  \langle \varphi ,  ( T_{L'} - W_{L'} -  z  )^{-1}  1_{L^{1/2+\alpha}  \Lambda_{1/2}^c  }   (W_{L'} - W_L )  \chi_L 
 ( T_{L} - W_{L} - z )^{-1}  \varphi \rangle_{L'}  |  \label{terminrhsest2}  \\
& \leq   \| \varphi \| \| ( T_{L'} - W_{L'} - z)^{-1} \|  \| 1_{L^{1/2+\alpha}  \Lambda_{1/2} }  (W_{L'} - W_L )   \|_\infty   \|  \chi_L  \|_\infty  \| ( T_{L'} - W_{L'} - z)^{-1} \| \| \varphi \|\nn  \\ 
& +\| \varphi \|  | {\rm Im} z|^{-1}    \|   (W_{L'} - W_L ) \chi_L \|_\infty  \| 1_{L^{1/2+\alpha}  \Lambda_{1/2}^c  } 
 ( T_{L} - W_{L} - z )^{-1}  \varphi  \| \nonumber \\
& \leq C  L^{-\epsilon}  | {\rm Im} z|^{-2}  \| \varphi \|^2  +   | {\rm Im} z|^{-1} \| \varphi\|^2 C L  C_0   \exp( - \kappa {L}^{\alpha})  \sup_{x  \in \supp \varphi} e^{   \langle x  \rangle  } 
 \to 0 , \nonumber 
\end{align} 
as $L \to \infty$. We note that 
for the decay of the term in Line  \eqref{terminrhsest1}  we used     Hypothesis \ref{hyp1} (ii) and for the 
decay of  the term in Line  \eqref{terminrhsest2}  we used   Lemma  \ref{combesthomas}
together with   Hypothesis \ref{hyp1} (i).

%Now this grows proportional to  $L$. % but the above term decays exponentially in $L$. 

\end{proof}

Now finally with Proposition  \ref{resulforcompactsupport} on our hand, we can turn to the proof of Theorem \ref{mainThm}.
\begin{proof}[Proof of Theorem   \ref{mainThm}]

In Proposition  \ref{resulforcompactsupport}, we have shown that the assertion of   Theorem \ref{mainThm}  holds provided  $\varphi$ 
vanishes outside of a sufficiently large ball. Thus we now assume  that $\varphi \in L^2(\R^d)$. 

 First,  we  choose $l > 0$   (independent of $L$) and  decompose   $\varphi = 1_{\Lambda_l} \varphi + 1_{\Lambda_l^c} \varphi$.
  Using first   linearity of the inner product and  the triangle inequality and  then   Cauchy-Schwarz inequality as well as  \eqref{ln-1} we   obtain
 %i.e.$\| ( T_L + W_L - z )^{-1} \| \leq |{\rm Im} z |^{-1}$ 
\begin{align}
 & | \langle  \varphi  , ( T_L + W_L - z )^{-1}  \varphi \rangle_L -   \langle 1_{\Lambda_{l}}   \varphi  , ( T_L + W_L - z )^{-1}  1_{\Lambda_{l}} \varphi \rangle_L  | \nn
\\
& \leq 
 | \langle    1_{\Lambda_{l}^c} \varphi , ( T_L + W_L - z )^{-1}  \varphi \rangle_L | +  | \langle  1_{\Lambda_{l}}\varphi , ( T_L + W_L - z )^{-1}  1_{\Lambda_{l}^c} \varphi \rangle_L  | \nn \\
 & \leq  |  {\rm Im} z|^{-1}  \|   1_{\Lambda_{l}^c} \varphi  \| ( \| \varphi \| + \|  1_{\Lambda_{l}}\varphi \|) .
 %\overset{l \to \infty}{\to} 0. 
\label{densitybound} 
\end{align} 
Observe that  by dominated convergence  $\|   1_{\Lambda_{l}^c} \varphi \| $, appearing 
on the RHS of \eqref{densitybound},  tends to zero as $l \to \infty$, independently of $L \geq 0$. %we use dominated convergence to conclude that   the LHS of  \ref{densitybound} tends to zero as well.
Let $\epsilon > 0$.  Thus  by \eqref{densitybound}   we can choose an  $l > 0 $ such that  for all $L \geq 1$ 
\begin{align} 
| \langle  \varphi  , ( T_L + W_L - z )^{-1}  \varphi \rangle_L -   \langle   1_{\Lambda_{l}}  \varphi,
 ( T_L + W_L - z )^{-1}  1_{\Lambda_{l}} \varphi \rangle_L  |  < \epsilon/3 . \label{densitybound2} \end{align}  
On the other hand  by aforementionend Proposition  \ref{resulforcompactsupport} 
%\tcr{.. converges in $L$ and therefore is a Chauchy sequence, which means that} 
we can  choose an $L_0 \geq 0 $ such that 
\begin{align}
| \langle    1_{\Lambda_{l}} \varphi  , ( T_L + W_L - z )^{-1}  1_{\Lambda_{l}} \varphi \rangle_L  - 
\langle   1_{\Lambda_{l}}  \varphi  , ( T_{L'} - W_{L'} - z )^{-1}  1_{\Lambda_{l}} \varphi \rangle_{L'} | < \epsilon/3  \label{densitybound3} 
\end{align} 
for all $L,L'  \geq L_0$. Thus it follows   for $L, L' \geq L_0$ by means of the triangle inequality from \eqref{densitybound2}  (used both for the $L$ as well as for the $L'$ summand) and  \eqref{densitybound3} that 
\begin{align*}
& | \langle   \varphi  , ( T_L + W_L - z )^{-1}   \varphi \rangle_L  - 
\langle      \varphi , ( T_{L'} - W_{L'} - z )^{-1} \varphi \rangle_{L'} |  \\
& \leq | \langle    1_{\Lambda_{l}} \varphi  , ( T_L + W_L - z )^{-1}  1_{\Lambda_{l}} \varphi \rangle_L  - 
\langle    1_{\Lambda_{l}} \varphi  , ( T_{L'} - W_{L'} - z )^{-1}  1_{\Lambda_{l}} \varphi \rangle_{L'} | \\
& + | \langle  \varphi  , ( T_{L'} - W_{L'} - z )^{-1}  \varphi \rangle_{L'} -   \langle   1_{\Lambda_{l}}  \varphi, ( T_{L'} - W_{L'} - z )^{-1}  1_{\Lambda_{l}} \varphi \rangle_{L'}  |  \\
& + | \langle  \varphi  , ( T_L + W_L - z )^{-1}  \varphi \rangle_L -   \langle  1_{\Lambda_{l}}  \varphi , ( T_L + W_L - z )^{-1}  1_{\Lambda_{l}} \varphi \rangle_L  |   \\
& < \frac{ \epsilon}{3} + \frac{ \epsilon}{3} +  \frac{ \epsilon}{3}  = \epsilon . 
\end{align*}
This shows \eqref{maintech1}, i.e.,  that \(
\langle  \varphi , ( T_L + W_L - z )^{-1} \varphi \rangle_L 
\) as a sequence in $L$ is a Cauchy sequence and  thereby converges.

Finally, observe that \eqref{maintech2}
now 
follows  from by  similar $\frac{\epsilon}{3}$-argument from \eqref{maintech1} using the  density of the functions $r_z : \R \to \C, x \mapsto (x-z)^{-1}$, for $z \in \C \setminus \R$,  in $C_\infty(\R)$. % := \{ f : \R \to \C : \text{continous and vanishing at infinity} \}$. 
\end{proof} 

\section{Linearly  bounded  potentials: spectral  measures} 

\label{specmeasgen}

In this  section  we will show 
 a generalized version of  Theorem  \ref{ThmFolgPMTgen-0}, which holds for a larger class of  linearly bounded 
 potentials.  We will show the vague convergence and then use a Portemanteau theorem version \ref{smallPMT} to conclude the weak convergence.

\begin{theorem} \label{ThmFolgPMTgen} 
Suppose Hypothesis \ref{hyp1} is satisfied  or more generally  the  assertion of Theorem  \ref{mainThm} holds.  Let $\varphi \in L^2(\R^d)$. 
Then there exists a unique measure $\mu_{\varphi}$ on $\R$ such that  for all $f \in C_c(\R)$ 
\begin{align}
\lim_{L \to \infty} \langle \varphi ,  f(T_L + W_L  ) \varphi \rangle =  \int f(\lambda) d\mu_{\varphi}(\lambda) .
\end{align} 
Furthermore, the spectral measures of $T_L +V_L$ with respect to $\varphi |_{\Lambda_L}$, denoted by   $\mu_{\varphi,L}$, 
converge   weakly   to the measure $\mu_{\varphi}$, i.e.  
\begin{align}
\label{weakKonv}
\int_\R f(\lambda)  d \mu_{\varphi}(\lambda)  = \lim_{L \to \infty} \int_\R  f(\lambda) d   \mu_{\varphi,L}(\lambda).
\end{align}
for all bounded continuous $f$ on $\R$.
In particular,  all of the statements from the  Portmanteau Theorem, i.e., Theorem  \ref{smallPMT},  %Appendix \ref{ApC} %\cite{Klenke.2014}   
 hold for the sequence $\mu_{\varphi,\omega,L}$.
\label{masskonv}
\end{theorem} 

\begin{remark}{\rm We show the existence of the limiting measure $\mu_\varphi$ by the Riesz-Markov theorem.
Alternatively  one could study  the 
 infinite volume  Hamiltonian directly, see [EYS], which can be shown to be self-adjoint [Kirsch].
We believe  that  the spectral  measures of the infinite volume Hamiltonian with respect to $\varphi$
 coincides  with   $\mu_\varphi$.
%
%\mu_\varphi$ is exactly the spectral measure of the infinite 
%volume Hamiltonian. 
%Furthermore, it is reasonable to ass that the limiting measure is equal to the spectral measure of  
% one could work with the spectral measure of the infinite volume 
%Hamiltonian. In that case, one would have to establish self-adjointness, which is not obvious 
%as that infinite volume Hamiltonian is unbounded from below cf. \cite{Kirsch},  and one would have to 
%establish convergence of operators in some sense, cf.  VIII.7 in \cite{RS1}. 
%\tcr{Robert sagt, dass er die Bemerkung nciht versteht, sich da aber auch nicht so auskennt}
}
\end{remark}

\begin{proof}
 By  the assertion of Theorem  \ref{mainThm}  we can define for any $f \in C_c(\R)$ 
\begin{align} \label{defofI-2} 
I_{\varphi}(f)  := \lim_{L \to \infty} \langle \varphi ,  f( T_L+ W_L ) \varphi \rangle_L  .
\end{align} 
Then  it follows from the spectral theorem that $I_{\varphi}$ defines a positive linear functional
on $C_c(\R)$. Thus by the Riesz-Markov theorem, there exists a unique measure $\mu_{\varphi}$ on $\R$
such that 
\begin{align}  \label{defofmu-2} 
I_{\varphi}(f)  = \int f(\lambda) d \mu_{\varphi}(\lambda) 
\end{align} 
for all $f$ in $C_c(\R)$. On the other hand 
the spectral measure $\mu_{\varphi,L}$ satisfies 
\begin{align}  \label{defofmuL-2} 
 \langle \varphi ,  f(T_L+ W_L) \varphi \rangle_L  = \int_\R  f(\lambda) d   \mu_{\varphi,L}(\lambda)
\end{align}
for all bounded Borel measurable functions $f$ on $\R$. As an immediate consequence of  \eqref{defofI-2}--\eqref{defofmuL-2} 
we obtain for all $f \in C_c(\R)$ 
\begin{align}
\lim_{L \to \infty}  \int_\R  f(\lambda) d   \mu_{\varphi,L}(\lambda) = \int f(\lambda) d \mu_{\varphi}(\lambda) 
\end{align} 
and hence  $\mu_{\varphi,L}$ converges vaguely to $\mu_{\varphi}$, cf.  Definition \ref{defofvagueweak}. 
%We will show the weak convergence by using the Portmanteau-like Theorem \ref{smallPMT}, which stats that  we can show vage convergence of the measures, i.e.\eqref{CcKonv}  for $f \in C_c(\R)$ instead of all bounded continuous functions as.   
%
%  On the other hand let $\mu_{\varphi,\omega,L}$ be the spectral measure of $H_L$.
%Then it follows immediately from the definitions that % \eqref{poslinfunc-1} that 
%\begin{align}
%\label{CcKonv}
%\int_\R f(\lambda)  d \mu_{\varphi,\omega}(\lambda) = \lim_{L \to \infty} \langle \varphi ,  f( H_L ) \varphi \rangle_L  = \lim_{L \to \infty} \int_\R  f(\lambda) d   \mu_{\varphi,\omega,L}(\lambda) ,
%\end{align} 
%where the last identity follows from the spectral theorem. 
%
%This shows the vage convergance of the measures.
To show weak convergence we want to  apply Theorem \ref{smallPMT}. For this,  we  consider the following using \eqref{defofmuL-2} 
\begin{align*}
\mu_{\varphi,L} ( \R ) 
= \int_{\R} 1  d  \mu_{\varphi,L}
= \langle \varphi ,  1 \varphi \rangle_L  =\int_{\R^d} 1_{\Lambda_L} |\varphi (x)|^2 d x.
\end{align*}
Without loss of generality we can assume that $\varphi$ satisfies $\| \varphi \|_2 \leq 1$ since otherwise we can normalize it and rescale it back afterwords.  Then $\mu_{\varphi,L} (\R)=\| \varphi \|_2 \leq 1$ holds and it is in $\mathcal{M}_{\leq 1}(\R)$.
Now $1_{\Lambda_L} |\varphi (x)|^2$ is  monotonically increasing  in $L$ and bounded by $|\varphi (x)|^2$ with $1_{\Lambda_L} \to 1$ for $L \to \infty$ which by monotone convergence theorem means
\begin{align*}
\lim_{L \to\infty }\mu_{\varphi,L} ( \R ) 
=\lim_{L \to\infty }\int_{\R^d} 1_{\Lambda_L} |\varphi (x)|^2 d x
=\int_{\R^d}  |\varphi (x)|^2 d x= \langle \varphi ,  1 \varphi \rangle
= \int_{\R} 1  d  \mu_{\varphi}
= \mu_{\varphi}(\R).
\end{align*}

Combing this result with the fact the the sequence $\mu_{\varphi,L}$ of measures  converges vaguely to  $\mu_{\varphi}$ 
and that $\R$ with standard topology is locally compact as well as Polish, see Definition \ref{defofpolish}, 
 shows that we can apply Theorem \ref{smallPMT}. Thus, we obtain weak convergence.
%\tcr{ALT:
%This,  together with the vague convergence (see the discussion after Theorem \ref{feynmankac-3}) as well as the fact that $\R$ with standard topology is locally compact as well as polish (i.e.  completely metrizable and separable) lets us apply Theorem \ref{smallPMT}, which grants the weak convergence and ends the proof.}
\end{proof}

\section{Proofs for the Poisson random potential} 

\label{secrandompotsatisfied} 

In this section    we show, in Lemma \ref{vldifflarge} below, that the  Poisson distributed random potential satisfies almost surely part (ii) of Hypothesis \ref{hyp1}. 
In the Appendix \ref{ApA} we will show  in  Lemma \ref{lemboundonpotbox}, that part (i) of Hypothesis \ref{hyp1} holds as well.
This will allow us   to prove Theorems   \ref{feynmankac-3} and  \ref{ThmFolgPMTgen-0} at the end of this section  by means of   Theorems  \ref{mainThm} and \ref{ThmFolgPMTgen}, respectively.

\begin{lemma} \label{vldifflarge}   Assume that   \eqref{propofprofileB}  i.e.
$
|B(x)| \leq C_B \langle x \rangle^{-d-1-\epsilon} 
$
 for some $\epsilon > 0$ holds.
%$$
%| B(x ) |  \leq C_B \langle x \rangle^{-d-1-\epsilon} 
%$$
%for some $\epsilon > 0$. 
Let $\alpha \in [0,1/2)$. Let $ O  = \Lambda_{1/2}$. Then for almost all 
 $\omega$ there exists a constant $C_\omega$  such that for all sufficiently large  $L \leq L'$ 
\begin{align}
\sup_{x \in \Lambda_{L}} |  1_{L^{1/2+\alpha} O}(x)   ( V_L(x) - V_{L'}(x)  ) |  \leq C_\omega L^{-\epsilon}  .
\end{align} 
\end{lemma}

\begin{proof} 
%For the proof we use  Lemma \ref{poissonpointsestforlarge}.
%{\tt  FIX modulo sum vs  integral corrections }{\tt  FIX $\mu_L$ versus $\mu$. ESY Section 3.3}  
By Lemma  \ref{poissonpointsestforlarge}  there exists  almost surely
a $Z >0$
such that 
\begin{align}\label{probestfromesyacta} 
 \int_{Q_k} d | \mu_\omega|(y) \leq  Z \langle k \rangle \text{ for all } k \in \Z^d ,
\end{align} 
where  $| \mu_\omega|$ denotes the total variation of the random measure $\mu_\omega$ and  we recall the definition of $Q_k$ given in \eqref{Qkdef} as the cube with side length 2 centered at $k$.   
%\begin{align}
%\omega \in \Omega_Z := \left\{ \omega : \int_{Q_k} d | \mu_\omega|(y) \leq  Z \langle k \rangle \text{ for all } k \in \Z^d \right\} .
%\end{align} 
Choose $L \geq 2$ sufficiently large that $L  \geq 2  L^{1/2+\alpha}$.
Let $x \in L^{1/2+\alpha} O$. Obviously, $x \in \Lambda_L$. 
Using  %\eqref{bumpnotationres}
\eqref{defofpot101}  -- the definition of $V_{L,\omega}(x)$ -- we find 
%\tcr{erstenschritt ausführlicher machen.. also einen zeischenschritt}
%\begin{equation}\label{bumpnotationres} 
% V_{L,\omega}(x) =  \sum_{\gamma=1}^M V_{L,\gamma}(x)   \quad \text{ with } \quad V_{L,\gamma}(x) := v_\gamma B_\#(x-y_{L,\gamma}) .
%\end{equation} 
%Using that $B_{\#,L}(x)= B_{\#,L'}(x)$ for $x \in \Lambda_L$ we find  
\begin{align}
&  |     V_{L'}(x) -   V_L(x)  |= \left| 
\int_{\Lambda_L} B_{\#,L}(x - y) d \mu_{\omega}(y) - \int_{\Lambda_{L'}} B_{\#,L'}(x - y) d \mu_{\omega}(y) \right| 
 \nn  \\
& 
 \leq   \left|  \int_{\Lambda_{L'} \setminus \Lambda_L} B_{\#,L'}(x-y)  d\mu_\omega(y) \right| +  \left| 
 \int_{  \Lambda_L} \left( B_{\#,L'}(x-y) - B_{\#,L}(x-y) \right)  d\mu_\omega(y) \right|  \label{diffest122}
\end{align} 

%\begin{equation}\label{bumpnotationres2} 
% V_{L,\omega}(x) =  \sum_{\gamma=1}^M V_{L,\gamma}(x)   \quad \text{ with } \quad V_{L,\gamma}(x) := v_\gamma B_\#(x-y_{L,\gamma}) .
%\end{equation} 
%\begin{align} \label{defofpot101} 
%V_{L,\omega}(x) := \int_{\Lambda_L} B_\#(x - y) d \mu_{L,\omega}(y),
%\end{align} 

To estimate the first term on the right hand side of \eqref{diffest122} we 
first observe that   for $z,y  \in \Lambda_{L'}$  we have $| z-y |_\infty \leq | z |_\infty + | y |_\infty \leq \frac{L'}{2} +  \frac{L'}{2}= L'$, which concludes 
\begin{align}
\label{inclusionLambda}
z - y \in \Lambda_{2L'} \subset \bigcup_{n \in \Z^d : |n|_\infty \leq 1 } \left( \Lambda_{L'} + n L' \right) ,
\end{align}
i.e.,  there is an $n_0 \in \Z^d$ with $|n_0 |_\infty \leq 1$ such that $z-y - n_0 L' \in \Lambda_{L'}$.
So from \eqref{inclusionLambda}
together with the periodic definition of the $B_{\#,L'}$ (see \eqref{defofperiodicext}), as well as    \eqref{propofprofileB}  we obtain 
\begin{align} \label{boundonBhashtag} 
|B_{\#,L'}(z-y) | &  =  |B_{\#,L'}(z-y - n_0 L') | =  |B(z-y - n_0 L') |   \nonumber   \\
& \leq     \max_{n \in \Z^d : |n|_\infty \leq 1 } |B (z - y - n L' ) | \leq  \max_{n \in \Z^d : |n|_\infty \leq 1 } C_B \langle z - y - n L' \rangle^{-d-1-\epsilon}  . 
\end{align}
For $a,b \in [-\frac{l}{2}, \frac{l}{2}]$ it is straight  forward 
to see by a distinction of cases, that  
\begin{align} \label{infmaxineqoned} 
\min_{n \in \{ 0,1,-1\} } | a - b - n l |  \geq \min \{|a - b|,  |a+b| \} .
\end{align} 
Thus using   monotonicity we see  from  \eqref{infmaxineqoned}   that 
\begin{align} \label{boundonBhashtag-20}  \nn 
& \max_{n \in \Z^d : |n|_\infty \leq 1 }\langle z - y - n L' \rangle^{-d-1-\epsilon}   =
\big( 1 + \min_{n \in \Z^d : |n|_\infty \leq 1 }\sum_{j=1}^d |z_j - y_j - n_j L' |^2   \big)^{ -(d+1+\epsilon)/2} \\ 
 & \leq   \big( 1 + \min_{\sigma \in \{-1,1\}^d }\sum_{j=1}^d | \sigma_j z_j - y_j |^2   \big)^{ -(d+1+\epsilon)/2} = \max_{\sigma \in \{-1,1\}^d } \langle \sigma z - y \rangle^{-d-1-\epsilon}  , 
\end{align}
where we denoted by   $\sigma z$ the vector in $\R^d$ with $j$-th component given by $\sigma_j z_j$. 
%Furthermore,  we  use that by  $\bigcup_{k \in \Z^d} Q_k = \R^d $ it follows for any subset $M$ of $\R^d$ 
%we have 
%\begin{align} \label{coveringofaset} 
%M \subset    \bigcup_{ k \in  \Z^d : Q_k \cap M \neq \emptyset  } Q_k . 
%\end{align}  
We find using   \eqref{boundonBhashtag}, \eqref{boundonBhashtag-20},  as well as the fact that $\mathbb{R}^d = \bigcup_{ k \in \Z^d }   Q_k $, which gives us the covering  $$ \Lambda_{L'} \setminus \Lambda_L \subset \bigcup_{\substack{ k \in \Z^d : \\ Q_k  \cap ( \Lambda_{L'} \setminus \Lambda_L)  \neq \emptyset}}   Q_k 
,
$$ 
 and    \eqref{probestfromesyacta} that 
%%\tcr{ist das eine Folgerung, doer eine Annahme?}
%%$x - y \in \bigcup_{n \in \Z^d : |n|_\infty \leq 1 } \left( \Lambda_{L'} + n L' \right) $  \tcr{etwas Intuition vermitteln? nochmal gemeinsam anschauen}
%and therefore
\begin{align}
&   \left|  \int_{\Lambda_{L'} \setminus \Lambda_L} B_{\#,L'}(x-y)  d\mu_\omega(y) \right| \nn \\
& \leq   \int_{\Lambda_{L'} \setminus \Lambda_L}  \max_{\sigma \in \{-1,1\}^d } C_B \langle \sigma  x - y  \rangle^{-d-1-\epsilon}    d | \mu_\omega|(y) \nn \\
& \leq  C_B   \sum_{\substack{ k \in \Z^d : \\ Q_k  \cap ( \Lambda_{L'} \setminus \Lambda_L)  \neq \emptyset}}  
\int_{Q_k} \max_{\sigma \in \{-1,1\}^d } \sup_{y \in Q_k}  \langle \sigma x - y  \rangle^{-d-1-\epsilon}    d | \mu_\omega|(y) \nn \\
& \leq  C_B 
 \sum_{\substack{ k \in \Z^d : \\ Q_k  \cap ( \Lambda_{L'} \setminus \Lambda_L)  \neq \emptyset}}  
 \max_{\sigma \in \{-1,1\}^d } \sup_{y \in Q_k}  \langle \sigma x - y  \rangle^{-d-1-\epsilon}
Z \langle k \rangle \nn \\
%& =  C_B  \sum_{\sigma \in \{-1,1\}^d }
% \sum_{\substack{ k \in \Z^d : \\ Q_k  \cap ( \R^d \setminus \Lambda_L)  \neq \emptyset}}  
% \sup_{y \in Q_k}  \langle \sigma x - y  \rangle^{-d-1-\epsilon}
%Z \langle k \rangle \nn \\
& \leq   C_B  \sum_{\sigma \in \{-1,1\}^d }
 \sum_{\substack{ k \in \Z^d \setminus \Lambda_{L-1}}}  
 \sup_{y \in Q_k}  \langle \sigma x - y  \rangle^{-d-1-\epsilon}
Z \langle k \rangle 
%\nn \\
%& =  C_B  \sum_{n \in \Z^d : |n|_\infty \leq 1 } 
% \sum_{\substack{ k \in \Z^d : \\ Q_k  \cap ( \Lambda_{L'} \setminus \Lambda_L)  \neq \emptyset}}  
% \sup_{y \in Q_k + n L'}  \langle x - y  \rangle^{-d-1-\epsilon} Z \langle k \rangle \nn \\
%& =   C_B  \sum_{n \in \Z^d : |n|_\infty \leq 1 }  \sum_{k \in \Z^d \cap ( \Lambda_{L'} \setminus \Lambda_L) + n L'}  
% \sup_{y \in Q_k}  \langle x - y \rangle^{-d-1-\epsilon} Z \langle k - n L' \rangle \nn \\
%& \leq   C_B  C_d   \sum_{k \in \Z^d  \setminus \Lambda_L }  
% \sup_{y \in Q_k}  \langle x - y \rangle^{-d-1-\epsilon} Z \langle k \rangle 
 \label{estin11} , 
\end{align} 
where  the last line follows from the inclusion of sets for $L \geq 2$
$$
\{ k \in \Z^d : Q_k  \cap ( \Lambda_{L'} \setminus \Lambda_L)  \neq \emptyset\} \subset  \Z^d \setminus \Lambda_{L-1} .
$$
%we introduced the notation   
% $C_d := \sum_{n \in \Z^d : |n|_\infty \leq 1 }$. 
Now the sum on the right hand side,  of    \eqref{estin11}, can be estimated   as follows in terms of  an integral 
by means of   Lemma \ref{lem:discrriem}, using monotonicity and the following inequality 
\begin{align} \label{boxestintsum}
 \sup_{ |\xi|_\infty \leq1/2}   \langle k + \xi  \rangle \leq (2d+1)^{1/2} \inf_{|\xi'|_\infty \leq1/2}  \langle k + \xi' \rangle , 
 \end{align} 
 which follows from
 $$\langle k + \xi  \rangle^2  = 1 + |k + \xi|^2 \leq 1 + 2 |k|^2 + 2 |\xi|^2  \leq 1 + 2 |k|^2 + 2 d  \leq ( 2 d +1 ) \langle k \rangle^2  , $$ 
 for all $ k , \xi  \in \R^d $ with $|\xi  |_\infty \leq 1$ .  Recall that we assumed
$x \in L^{1/2+\alpha} O$.   We  conclude by Lemma \ref{lem:discrriem}
that for some finite constant denoted by ${\rm const.}$  independent of $L$, 
which may change from line to line, 
making repeated use of the elementary inequalities \eqref{chineseest-1}  and \eqref{chineseest-2} and that for $k \in \R^d \setminus \Lambda_{L-2} $ we have $|k|_\infty \geq \frac{L}{2} - 1 $
and estimating  the  integral in terms of   polar coordinates in the second to last step,
we obtain for $L \geq 3$ 
\begin{align}
%&  |  1_{L^{1/2+\alpha} O}(x)   ( V_L(x) - V_{L'}(x)  ) | \\
%&  =  |  1_{L^{1/2+\alpha} O}(x)   \sum_{\gamma=1}^{M'} 1_{y_{L',\gamma} \notin \Lambda_L}  v_\gamma B_{\#,L'}(x-y_{L',\gamma})  | \\
%&  \leq C_B  \sum_{k \in \Z^d : k  \notin \Lambda_{L/2} } |x - k |^{-d-1-\epsilon} c_v Z \langle k \rangle \\
  \eqref{estin11}
& \leq  {\rm const.}  \sum_{\sigma \in \{-1,1\}^d }\int_{ \R^d \setminus \Lambda_{L-2} } \langle    k  - \sigma x \rangle^{-d-1-\epsilon}   \langle k  \rangle  dk  \nn \\
& \leq  {\rm const.}  \sum_{\sigma \in \{-1,1\}^d }\int_{ \R^d \setminus \Lambda_{L-2} } (1 +  |  k  - \sigma x|_\infty)^{-d-1-\epsilon}   \langle k  \rangle  dk  \nn \\
& \leq  {\rm const.}  \sum_{\sigma \in \{-1,1\}^d }\int_{ \R^d \setminus \Lambda_{L-2} } (1 +  |  k |_\infty  - | \sigma x|_\infty)^{-d-1-\epsilon}   \langle k  \rangle  dk  \nn \\
& \leq  {\rm const.}  \sum_{\sigma \in \{-1,1\}^d }\int_{ \R^d \setminus \Lambda_{L-2} } \left(1 +  \frac{|  k |_\infty}{2}   +   \frac{L}{4} -\frac{1}{2} - \frac{L^{1/2+\alpha}}{4} \right)^{-d-1-\epsilon}   \langle k  \rangle  dk  . \label{estintrhs}  %\nn \\
%& \leq  {\rm const.}  \int_{ |k|_\infty  \geq L-2 } |k|^{-d-1-\epsilon}  |k|  dk  \nn \\
%& \leq  {\rm const.}  \int_{L-2    }^\infty r^{-d-1-\epsilon}  r r^{d-1} dr = 
%{\rm const.}  \int_{L-2    }^\infty r^{-1-\epsilon}  dr  \nn \\
%%& \leq  {\rm const.} \int_{r \geq L/2-1} |r - L^{1/2+\alpha} |^{-d-1-\epsilon}   r |S_{d-1}| r^{d-1}  dr \nn \\
%%& \leq  {\rm const.}     \int_{r \geq L/2-1} |r/2 + L/4 - 1/2  - L^{1/2+\alpha} |^{-d-1-\epsilon}  r^d   dr \nn \\
%%& \leq   {\rm const.}  \int_{r \geq L/2-1}  r^{-d-1-\epsilon}  r^{d} dr  \nn  \\
%%& \leq    {\rm const.}   \int_{r \geq L/2-1}  r^{-1-\epsilon}  dr  \nn \\
%& \leq  {\rm const.} \  L^{-\epsilon} ,  \label{diffest122+1}
%%\frac{(L/2)^{-\epsilon}}{\epsilon}.
\end{align} 
Estimating  the  integral  on the right hand side in \eqref{estintrhs}  in terms of   polar coordinates 
we obtain for $L \geq 3$ 
\begin{align}
%&  |  1_{L^{1/2+\alpha} O}(x)   ( V_L(x) - V_{L'}(x)  ) | \\
%&  =  |  1_{L^{1/2+\alpha} O}(x)   \sum_{\gamma=1}^{M'} 1_{y_{L',\gamma} \notin \Lambda_L}  v_\gamma B_{\#,L'}(x-y_{L',\gamma})  | \\
%&  \leq C_B  \sum_{k \in \Z^d : k  \notin \Lambda_{L/2} } |x - k |^{-d-1-\epsilon} c_v Z \langle k \rangle \\
  \eqref{estin11}
%& \leq  {\rm const.}  \sum_{\sigma \in \{-1,1\}^d }\int_{ \R^d \setminus \Lambda_{L-2} } \langle    k  - \sigma x \rangle^{-d-1-\epsilon}   \langle k  \rangle  dk  \nn \\
%& \leq  {\rm const.}  \sum_{\sigma \in \{-1,1\}^d }\int_{ \R^d \setminus \Lambda_{L-2} } (1 +  |  k  - \sigma x|_\infty)^{-d-1-\epsilon}   \langle k  \rangle  dk  \nn \\
%& \leq  {\rm const.}  \sum_{\sigma \in \{-1,1\}^d }\int_{ \R^d \setminus \Lambda_{L-2} } (1 +  |  k |_\infty  - | \sigma x|_\infty)^{-d-1-\epsilon}   \langle k  \rangle  dk  \nn \\
%& \leq  {\rm const.}  \sum_{\sigma \in \{-1,1\}^d }\int_{ \R^d \setminus \Lambda_{L-2} } \left(1 +  \frac{|  k |_\infty}{2}   +   \frac{L}{4} -\frac{1}{2} - \frac{L^{1/2+\alpha}}{4} \right)^{-d-1-\epsilon}   \langle k  \rangle  dk  \nn \\
& \leq  {\rm const.}  \int_{ |k|_\infty  \geq L-2 } |k|^{-d-1-\epsilon}  |k|  dk  \nn \\
& \leq  {\rm const.}  \int_{L-2    }^\infty r^{-d-1-\epsilon}  r r^{d-1} dr = 
{\rm const.}  \int_{L-2    }^\infty r^{-1-\epsilon}  dr  \nn \\
%& \leq  {\rm const.} \int_{r \geq L/2-1} |r - L^{1/2+\alpha} |^{-d-1-\epsilon}   r |S_{d-1}| r^{d-1}  dr \nn \\
%& \leq  {\rm const.}     \int_{r \geq L/2-1} |r/2 + L/4 - 1/2  - L^{1/2+\alpha} |^{-d-1-\epsilon}  r^d   dr \nn \\
%& \leq   {\rm const.}  \int_{r \geq L/2-1}  r^{-d-1-\epsilon}  r^{d} dr  \nn  \\
%& \leq    {\rm const.}   \int_{r \geq L/2-1}  r^{-1-\epsilon}  dr  \nn \\
& \leq  {\rm const.} \  L^{-\epsilon} .  \label{diffest122+1}
%\frac{(L/2)^{-\epsilon}}{\epsilon}.
\end{align} 
%where $|S_{d-1}|$ denotes the volume of $S_{d-1} := \{ x \in \R^d : |x|=1\}$.
Next we estimate the second term on the right hand side of \eqref{diffest122}. Using that by definition of the periodic extension, $B_{\#,L}(z)= B_{\#,L'}(z)$ for $z \in \Lambda_L$, and so  
\begin{align}
&   \left| 
 \int_{  \Lambda_L} \left( B_{\#,L'}(y-x) - B_{\#,L}(y-x) \right)  d\mu_\omega(y) \right| \nn \\
& =  \left| 
 \int_{ y \in   \Lambda_L: x-y \notin \Lambda_L} \left( B_{\#,L'}(y-x) - B_{\#,L}(y-x) \right)  d\mu_\omega(y) \right| . \label{estonrandompot2}
\end{align}  
Recall that by assumption    $x \in  L^{1/2+\alpha} O$. Suppose  $y \in \Lambda_L$ and  $x - y  \notin  \Lambda_L$. 
Then  $x - y -  n_0  L \in \Lambda_L$ for some $n_0 \in \Z^d$ with $|n_0|_\infty = 1$, it  follows  by  the triangle inequality 
\begin{align} \label{esonboxes1} 
|x - y -  n_0 L|_\infty \geq  L - | x|_\infty - |y|_\infty \geq L- \frac{1}{4}L^{1/2+\alpha}-L/2  \geq L/4 ,  
\end{align} 
and so by \eqref{propofprofileB} 
\begin{align} \label{estonprofile1111-1} 
|B_{\#,L}(y-x) |  = |B(y-x-n_0 L) | \leq C_B \langle x - y - n_0 L \rangle^{-d-1-\epsilon} \leq  
C_B (L/4)^{-d-1-\epsilon} .
\end{align} 
On the other hand   $x - y  - n_0 L' \in \Lambda_{L'}$  for some $n_0 \in \Z^d$.
If $n_0 \neq 0$  it follows  analogously as in  \eqref{esonboxes1} that  
\begin{align} \label{estondecaybound1} 
|x - y -  n_0 L'|_\infty \geq  L' - | x|_\infty - |y|_\infty  %\geq    L-L/2 - \frac{1}{4}L^{1/2+\alpha}
\geq    L - | x|_\infty - |y|_\infty \geq L/4  ,
\end{align} 
and if $n_0=0$ it follows (since by assumption $x-y \notin \Lambda_L$)  that 
\begin{align} \label{estondecaybound2} 
|x - y -  n_0 L'|_\infty =  |y-x|_\infty     \geq L/2 . 
\end{align} 
Hence  using \eqref{propofprofileB} and  inserting    \eqref{estondecaybound1}  and    \eqref{estondecaybound2}, respectively, we obtain 
\begin{align} \label{estonprofile2222-2} 
|B_{\#,L'}(y-x) |  = |B(y-x-n_0 L') | \leq C_B \langle x - y - n_0 L' \rangle^{-d-1-\epsilon} \leq  
C_B (L/4)^{-d-1-\epsilon} .
\end{align} 
Thus inserting \eqref{estonprofile1111-1} and  \eqref{estonprofile2222-2}  to estimate \eqref{estonrandompot2},
 we find using \eqref{probestfromesyacta}
\begin{align}
&  \left| 
 \int_{ y \in   \Lambda_L: x-y \notin \Lambda_L} \left( B_{\#,L'}(y-x) - B_{\#,L}(y-x) \right)  d\mu_\omega(y) \right| \nn \\
& \leq 2 C_B (L/4)^{-d-1-\epsilon}  \int_{ y \in   \Lambda_L: x-y \notin \Lambda_L}  d |\mu_\omega|(y)\nn  \\
%& \leq 2 C_B (L/4)^{-d-1-\epsilon}  \int_{ y \in   \Lambda_L}  d |\mu_\omega|(y) \nn \\
& \leq 2 C_B (L/4)^{-d-1-\epsilon} \sum_{k \in \Z^d \cap \Lambda_{L+1} }  \int_{ Q_k }  d |\mu_\omega|(y) \nn \\
& \leq 2 C_B (L/4)^{-d-1-\epsilon} \sum_{k \in \Z^d \cap \Lambda_{L+1} }  Z \langle k \rangle  \label{sumasintegralest} ,
\end{align}  
where in the third line we made use of the inclusion of sets $$\{  y \in   \Lambda_L: x-y \notin \Lambda_L\} \subset \Lambda_L \subset \bigcup_{k \in \Z^d \cap \Lambda_{L+1} } Q_k. $$ 
Now the sum  in   \eqref{sumasintegralest}   can be bounded 
in terms of the following  integral 
\begin{align} \label{estonsumofkchin}
& \sum_{k \in \Z^d \cap \Lambda_{L+1} }  \langle k \rangle 
\leq {\rm const.}  \int_{ \Lambda_{L+2}}  \langle k \rangle dk  \leq {\rm const.} \, L^{d+1}  , 
\end{align} 
where we used again  Lemma \ref{lem:discrriem} together with the estimate  \eqref{boxestintsum}. 
Thus inserting \eqref{estonsumofkchin} into  \eqref{sumasintegralest} and using \eqref{estonrandompot2} we arrive at
\begin{align}
&   \left| 
 \int_{  \Lambda_L} \left( B_{\#,L'}(y-x) - B_{\#,L}(y-x) \right)  d\mu_\omega(y) \right|  
 \leq {\rm const.} L^{-\epsilon}  . \label{diffest122+2}
\end{align}  
Finally,  inserting   \eqref{diffest122+1} and  \eqref{diffest122+2} into \eqref{diffest122}
 yields the claim. 
\end{proof}

\label{proofofmainpoisson} 
\begin{proof}[Proof of  Theorem \ref{feynmankac-3} ]

This follows immediately  from Theorem \ref{mainThm} and the fact that Hypothesis \ref{hyp1},  in particular Parts (i) and (ii)
are  satisfied almost surely by Lemmas \ref{vldifflarge}  and \ref{lemboundonpotbox}, respectively.
\end{proof} 
\begin{proof}[Proof of  Theorem \ref{ThmFolgPMTgen-0} ]

This follows immediately  from Theorem \ref{ThmFolgPMTgen} and the fact that Hypothesis \ref{hyp1},  in particular Parts (i) and (ii)
are  satisfied almost surely by Lemmas \ref{vldifflarge}  and \ref{lemboundonpotbox}, respectively.
\end{proof}

\appendix 

\section{Probabilistic Estimates on the Random Potential} 
\label{ApA}
The probabilistic results in this appendix are used in Section \ref{proofs}.
% To obtain estimates which are uniform in $L$, we introduce the following infinite volume measure as in  \cite{ErdosSalmhoferYau.2008a}. 
 Note that Lemma \ref{poissonpointsestforlarge} as well as its proof  is from \cite{ErdosSalmhoferYau.2008a}. 
%Let  $\mu_\omega$ be a Poisson point process on $\R^d$ with homogeneous unit density and 
%with independent, identically distributed random masses.
%More explicitly, for almost all realizations $\omega$, it consists of a countable, locally finite 
%collection of points $\{ y_\gamma(\omega) \in \R^d : \gamma \in \N \}$, and random 
%weights $\{ v_\gamma(\omega) \in \R : \gamma \in \N \}$ such that 
%\begin{align}
%\mu_\omega = \sum_{\gamma=1}^\infty v_\gamma \delta_{y_\gamma(\omega )} 
%\end{align} 
%where $\delta_a$ denotes the Dirac mass at $a \in \R^d$. 
%The Poisson process $\{ y_\gamma(\omega) \}$ is independent of
%the weights $\{ v_\gamma(\omega) \}$. The weights are real iid
%random variables   with distribution $\boldsymbol{P}_v$ and
%with moments $m_k :=  {\bf E}_v v_\gamma^k $ satisfying
%\begin{align} \label{assmoment} 
%m_2 = 1 , \quad  m_{2d } < \infty .
%\end{align} 
%
%\begin{remark} If one restricts the above measure to $\Lambda_L$, then one 
%obtains the measure $\mu_{\omega,L}$. 
%\end{remark} 
To state the next lemma we define 
  the cube  \begin{align} Q_k := \{ y \in \R^d :  |y - k |_\infty \leq 1\} \label{Qkdef}  \end{align}   centered at $k \in \R^d$ with side length two. This lemma will be used in the proof of subsequent Lemma \ref{lemboundonpotbox}

\begin{lemma} \label{poissonpointsestforlarge}  Define the following event for any $Z > 0$ 
\begin{align}
\Omega_Z := \left\{ \omega : \int_{Q_k} d | \mu_\omega|(y) \leq  Z \langle k \rangle \text{ for all } k \in \Z^d \right\} ,
\end{align} 
where $|\mu_\omega|$ denotes the total variation of the random measure $\mu_\omega$. 
Then $\lim_{Z \to \infty} P(\Omega_Z ) = 1 $.
\end{lemma} 
\begin{proof} For any fixed $k \in \Z^d$, let $N_k$ be the number of Poisson points in the cube $Q_k$.
Define for $k \in \Z^d$  the random variable 
$$
X_k(\omega)  :=  \int_{Q_k} d | \mu_\omega|(y). 
$$
We compute using the Hoelder inequality 
\begin{align*}
{\bf E}  X_k^{d+1} 
& = {\bf E}  \left( \int_{Q_k} d | \mu_\omega|(y) \right)^{d+1} =
 {\bf E}  \left( \sum_{\gamma=1}^\infty  |v_\gamma| 1_{y_\gamma \in Q_k}   \right)^{d+1} \\
 & ={\bf E}  \bigg( \sum_{\gamma_1, \cdots , \gamma_{d+1} = 1}^{N_k}  |v_{\gamma_1}|  \cdots |v_{\gamma_{d+1}}| \bigg)  \leq  {\bf E} \bigg(
 \sum_{\gamma_1, \cdots , \gamma_{d+1} = 1}^{N_k} \bigg)  {\bf E}_v ( |v_1 |^{d+1}  ) 
 %\int_{ |y-k| \leq 1} d | \mu_\omega |(y) \right\|^{d+1} 
\leq {\bf E} N_k^{d+1} | v_1|^{d+1} \leq C_d , 
\end{align*} 
using  \eqref{2.4d} and the fact that $N_k$ is  Poisson random variable with expectation ${\bf E}N_k$,
which is the volume of $Q_k$ (which is $2^d$). 
%Let 
%$$
% C_{Z,k} := \left\{ \omega : \int_{|y - k |_\infty \leq 1} d | \mu_\omega|(y)  >   Z \langle k \rangle \right\}
%$$
By Markov inequality 
\begin{align*}
P(  X_k > Z \langle k \rangle  ) = P \left(  X_k^{d+1} > ( Z  \langle k \rangle)^{d+1}   \right)  \leq \frac{ {\bf E}(X_k^{d+1}   )}{ Z^{d+1} \langle k \rangle^{d+1}  } \leq 
\frac{C_d }{ Z^{d+1} \langle k \rangle^{d+1}  }
\end{align*} 
and so 
\begin{align*}
P(\Omega_Z^c )  = P \left(\bigcup_{k \in \Z^d} \{ \omega : X_k > Z \langle k \rangle \}  \right) \leq \sum_{k \in \Z^d}
P(   X_k > Z \langle k \rangle   ) 
\leq \sum_{k \in \Z^d} \frac{C_d}{Z^{d+1} \langle k \rangle^{d+1}} = O(Z^{-d-1} ) ,
\end{align*} 
thus we have 
\begin{align*}
\lim_{Z \to \infty} {\bf P}(\Omega_Z) = 1 .
\end{align*}  
\end{proof}

Next we show a lemma which will be used in the proof of Lemma \ref{combesthomas}.
We recall the assumption in \eqref{propofprofileB}  that for some  $C_B$ and  $\epsilon > 0$  
%\begin{align}  \label{propofprofileB} 
%|B(x)| \leq C_B \langle x \rangle^{-d-1-\epsilon} 
%\end{align} 
%\tcr{Überleitung?}
\begin{align}  \label{propofprofileB-2} 
|B(x)| \leq C_B \langle x \rangle^{-d-1-\epsilon} .
\end{align}

\begin{lemma} \label{lemboundonpotbox}   For almost every $\omega$ there exists a constant $C$ such that for all $x \in \Lambda_L$ 
\begin{align}
| V_L(x) | \leq C \langle x \rangle   . 
\end{align} 
\end{lemma} 

Before we give the proof we recall the following elementary  identities 
\begin{align} \label{chineseest-1} |x|  + 1  \geq 
\langle x \rangle \geq 2^{-1/2}(|x|  + 1 ) ,
\end{align} 
\begin{align} \label{chineseest-2} 
   \sqrt{d}  |x|_\infty \geq  |x|  \geq |x|_\infty   .
\end{align}

The assumption   \eqref{propofprofileB}  implies for the periodic extension for $x \in \R^d$ we have 
\begin{align} \label{boundbhashtag}
|B_{\#,L}(x) |  %\leq \sup_{n \in \Z^d} C_B \langle  x - n L \rangle^{-d-1-\epsilon} =
\leq  C_B \langle  x - n_0 L \rangle^{-d-1-\epsilon}, 
\end{align} 
where $n_0 \in \Z^d$ is such that $x \in \Lambda_L + n_0 L$. 
Thus if $x \in \Lambda_L + n_0 L$ for some $n_0 \in \Z^d$  with  $| n_0 |_\infty \leq  1$
we find 
\begin{align} \label{boundbhashtag-2}
|B_{\#,L}(x) |  %\leq \sup_{n \in \Z^d} C_B \langle  x - n L \rangle^{-d-1-\epsilon} =
\leq   \sum_{n \in \Z^d : |n|_\infty \leq  1} C_B \langle  x - n L \rangle^{-d-1-\epsilon} .
\end{align} 

\begin{proof}
By Lemma \ref{poissonpointsestforlarge} there exists  almost surely 
 a constant $Z$ (depending on $\omega$) such that 
 \begin{align} \label{assumpinapp3}
 \int_{|y - k |_\infty \leq 1} d | \mu_\omega|(y) \leq  Z \langle k \rangle . 
 \end{align} 
Using this, we estimate   for $x \in \Lambda_L$  using \eqref{boundbhashtag-2} 
 \begin{align}
 | V_L(x) |  &  = \left| \int_{\Lambda_L}  B_\#(x-y) d\mu_\omega(y)  \right| \nn \\
 & \leq   \int_{\Lambda_L}   | B_\#(x-y)|  d | \mu_\omega(y)  | \nn \\
  & \leq  \sum_{k \in \Z^d \cap \Lambda_{L}}   \int_{Q_k}   | B_\#(x-y)|  d | \mu_\omega(y)  | \nn  \\
        & \leq   \sum_{k \in \Z^d \cap \Lambda_{L}}     
 \sup_{y  \in Q_k } | B_\#(x-y)|  Z  \langle  k \rangle \nn \\ %   \label{assumpinapp3-0}  \\
        & \leq  \sum_{k \in \Z^d \cap \Lambda_{L}}     
 \sup_{y  \in Q_k} \sum_{n \in \Z^d : |n|_\infty \leq  1} C_B \langle  x -y - n L \rangle^{-d-1-\epsilon}  Z  \langle  k \rangle  \label{unglweiter}.
\end{align}  We will now continue with a shift of variables,  in first step and by monotonicity, we can use
 $$\langle  k - n L \rangle  \leq  \langle  k  \rangle $$
if $k \in  \Z^d \cap \Lambda_{L} + n L$ in the second step. Finally in the third step we  introduce the notation $C_d :=  | \{n \in \Z^d : |n|_\infty \leq  1\}|$ and use by means of   \eqref{chineseest-1}  and  \eqref{chineseest-2} 
\begin{align}
 \inf_{y \in Q_k} \langle x - y \rangle &  \geq  2^{-1/2} \inf_{y \in Q_k} ( | x - y |_\infty + 1    ) \nn \\
  & \geq  2^{-3/2} \inf_{y \in Q_k} ( | x - y |_\infty + 2    ) \nn \\
  &  \geq   2^{-3/2} \inf_{y \in Q_k} ( | x - k |_\infty  - | k -  y |_\infty +2     )  \nn \\
   &  \geq   2^{-3/2} (| x - k |_\infty +1 ) .  \label{potestinf4}
\end{align}  
to continue at \eqref{unglweiter}
\begin{align}
  \eqref{unglweiter} &  \leq  C_B\sum_{n \in \Z^d : |n|_\infty \leq  1}\sum_{k \in \Z^d \cap \Lambda_{L} + n L}     
 \sup_{y  \in Q_k }   \langle  x -y\rangle^{-d-1-\epsilon}  Z  \langle  k - n L \rangle   \nn  \\     
  &  \leq    C_B\sum_{n \in \Z^d : |n|_\infty \leq  1}\sum_{k \in \Z^d \cap \Lambda_{L} + n L}     
 \sup_{y  \in Q_k}   \langle  x -y\rangle^{-d-1-\epsilon}  Z  \langle  k  \rangle  \nn   \\
  &  \leq    C_B C_d \sum_{k \in \Z^d }     
 \sup_{y  \in Q_k}   \langle  x -y\rangle^{-d-1-\epsilon}  Z  \langle  k  \rangle.  \label{potestinf3} 
 \end{align}   
Thus  using  \eqref{potestinf4}  and monotonicity in   \eqref{potestinf3}, as well as  $  \langle  k \rangle \leq (\sqrt{d} |k|_\infty + 1 )$ (cf. \eqref{chineseest-1},  \eqref{chineseest-2}) we estimate 
 \begin{align}
 | V_L(x) |  
         & \leq  C_B C_d   \sum_{k \in \Z^d }    (  2^{-3/2} (| x - k |_\infty +1 ))^{-d-1-\epsilon}  Z  \langle  k \rangle   \nn   \\
                   & \leq  C_B  C_d Z 2^{\frac{3}{2}(d+1+\epsilon)}\sum_{k \in \Z^d}      (  (| x - k |_\infty +1 ))^{-d-1-\epsilon}  (\sqrt{d} |k|_\infty + 1 )   \nn \\
                   & \leq  \sqrt{d} C_B C_d Z 2^{\frac{3}{2}(d+1+\epsilon)}\sum_{k \in \Z^d}    (  | x - k |_\infty +1 )^{-d-1-\epsilon}    \underbrace{(|k-x|_\infty + 1  + |x|_\infty )}_{\leq (|k-x|_\infty + 1)  (1+|x|_\infty )} \nn    \\
                   & \leq  \sqrt{d} C_B C_d Z 2^{\frac{3}{2}(d+1+\epsilon)} \sum_{k \in \Z^d}    (  | x - k |_\infty +1 )^{-d-\epsilon}    (1  + |x|_\infty ).   \label{EstWeiter2}
\end{align}
We are now going to use a shift of the summation variables and for $x\in \R^d$ we denote by $x'$ the unique element in   $ \Lambda_1$ such that  $x= k+x'$ for some $k\in \Z^d$ 
\begin{align*}                                
            \eqref{EstWeiter2}       & \leq  \sqrt{d} C_B C_d Z 2^{\frac{3}{2}(d+1+\epsilon)} \sum_{k \in \Z^d}   (  | x' - k |_\infty +1 )^{-d-\epsilon}    (1  + |x|_\infty )   \\
                   & \leq  \sqrt{d} C_B C_d Z 2^{\frac{3}{2}(d+1+\epsilon)} (1  + |x|_\infty ) \sum_{k \in \Z^d}   (  | k |_\infty - \underbrace{|x'|_\infty}_{\leq 1/2}+1 )^{-d-\epsilon}       \\
                     & \leq  \sqrt{d} C_B C_d Z 2^{\frac{3}{2}(d+1+\epsilon)} \sqrt{2} \langle x \rangle \sum_{k \in \Z^d}   \left(  | k |_\infty +\frac{1}{2} \right)^{-d-\epsilon}    .
 \end{align*} 

Since the sum is finite by well known estimates, the claim follows. 
\end{proof}

\section{Estimates on Integrals} 
\label{ApB}

In this Appendix, we present a collection of estimates on integrals as well as on discrete integrals, i.e.,  infinite sums.
Let us first mention a basic inequality.  For any $z \in \C$ we have the elementary bound 
\begin{equation}  \label{ln-12}
   | z^{-1}|   \leq   \left|   {\rm Im} z \right|^{-1}  .
\end{equation} 
If  $A$ is a selfadjoint operator then it follows from the spectral theorem and   \eqref{ln-12} that for any $z \in \C \setminus \R$ we have 
\begin{align}  \label{ln-1}
 \| (A - z)^{-1} \| \leq \frac{1}{|{\rm Im } z |} .
\end{align}

%The following lemma is a simple estimate,  which is applied in the proof of Lemma \ref{lemestfourdisc}.
%\begin{lemma}
%\label{ChSymbEst}
%For  $a,b \in \R$ with $a \leq b$ and  $ \langle \cdot \rangle^2 f \in L^\infty$ with $\langle \cdot \rangle$ defined in \eqref{Klammern}.  Then  $f \in L^1(\R)$ and
% \begin{align*}
%\int_{a}^b  \left|   f(x) \right|   dx \leq \pi \sup_{x \in \R} | \langle x \rangle^2 f(x) | .
%\end{align*}
%\end{lemma}
%\begin{proof}
%Using an estimate on the $\arctan$ and a straight forward calculation shows
% \begin{align*} 
%\int_{a}^b  \left|   f(x) \right|   dx
% &\leq    \int_{a}^{b} \langle x \rangle^{-2}\langle x \rangle^{2}  |  f(x) |  dx \nn \\
%  &\leq  \sup_{x \in \R} | \langle x \rangle^2 f(x) |   \int_{a}^{b}\frac{1}{1+x^2} dx \nn \\
%&=  \sup_{x \in \R} | \langle x \rangle^2 f(x) | \underbrace{\left[ \arctan(x) \right]_a^b}_{\leq \pi} \nn \\
% & \leq \pi \sup_{x \in \R} | \langle x \rangle^2 f(x) | < \infty .
%\end{align*}
%\end{proof}
%Moreover we shall need to estimate tracial expressions of resolvents. 
%Let us now  estimate a  finite volume expression. For this we shall make use of the following elementary lemma about Riemann integrals. 

\begin{lemma}\label{lem:discrriem}  Let $I  = [-c,c]^d$ and $I_j$, $j=1,...,N$, be   a partition (up to boundaries) of $I$  into translates of a cube  $Q$ which is centered at the 
origin. Let $\xi_j  \in  I_j$
and $ \Delta x_j = | I_j|$. Suppose $g \geq 0$ is a Riemann integrable function on $I$  and we have 
\begin{equation} \label{eq:estonint} 
\sup_{ \xi \in I_j} | f(\xi) |  \leq \inf_{ \xi \in I_j} g(\xi)  . 
\end{equation} 
Then 
\begin{align*} 
  \left|  \sum_{i=1}^N f(\xi_j) \Delta x_j  \right| \leq  \int_I g(x) dx  .
\end{align*} 
\end{lemma}
\begin{proof}  The statement follows directly from the theory of Riemann integration. 
\end{proof}

\section{Portemanteau Theorem} 
\label{ApC}
\begin{definition}
 \label{defofvagueweak} 
Let $E$ be a complete metric space equipped with the Borel-$\sigma$-algebra $\mathcal{E}$.  %Let $\mu ,  \mu_n$ for $n \in \N$ be finite Borel-measures on $(E,\mathcal{E})$.  
\begin{itemize}
\item[a)]  Let $\mu ,  \mu_n$ for $n \in \N$ be  finite measures on $(E,\mathcal{E})$.  
We say the   sequence of measures $(\mu_n)$  converges {\bf weakly}   to $\mu$ if
\begin{align}
\label{wvMconv}
\int_E f(\lambda)  d \mu(\lambda)  = \lim_{n \to \infty} \int_E  f(\lambda) d   \mu_n (\lambda)
\end{align}
holds for all bounded continuous functions $f$ on $E$. In that case we write
\begin{align*}
\mu = \text{w-}\lim_{n \to \infty} \mu_n.
\end{align*}
\item[b)] Let $\mu ,  \mu_n$ for $n \in \N$ be  Radon  measures on $(E,\mathcal{E})$.  
We say 
the sequence of measures $(\mu_n)$  converges {\bf vaguely}    to $\mu$ if \eqref{wvMconv} if  for all $f \in C_c(E)$. In that case we write
\begin{align*}
\mu = \text{v-}\lim_{n \to \infty} \mu_n.
\end{align*}
\end{itemize}

\end{definition}

\begin{definition}\label{defofpolish} 
A topological space \( (E,  \mathcal{T}) \) is called a   {\bf Polish}  space if it is separable and there exists a metric $d$ on $E$, which induces the topology $\mathcal{T}$, such that $(E,d)$ is a complete metric space.
\end{definition}
For  $E$  a toplogical space equipped with the Borel $\sigma$-algebra   $\mathcal{E}$,   we define a set of measures on $(E,\mathcal{E})$ by 
\begin{align}
\mathcal{M}_{\leq 1}(E):= \{ \mu \text{ is finite on $(E,\mathcal{E})$ with } \mu (E) \leq 1\}.
\end{align}
We note that on  a Polish space every  bounded measure is a Radon measure, cf.  \cite[ Theorem 13.6]{Klenke.2014}. 
The following theorem is a shorter version of the Portemanteau \cite[Theorem 13.16]{Klenke.2014}  in which we only state the equivalence of weak and vague measure convergence under certain conditions. Note that the original Portemanteau Theorem contains way more equivalent statements.
\begin{theorem}
\label{smallPMT}
Let \(E\) be a metric space that is  locally compact and Polish,  and let \(\mu, \mu_1, \mu_2, \ldots \in \mathcal{M}_{\leq 1}(E)\). Then the following statements are equivalent
\begin{itemize}
    \item[(i)] $\mu = \text{w-}\lim_{n \to \infty} \mu_n $.
    \item[(ii)]$\mu = \text{v-}\lim_{n \to \infty} \mu_n $ and \(\mu(E) = \lim_{n \to \infty} \mu_n(E)\).
\end{itemize}
\end{theorem}

\section{Acknowledgements}
We thank Robert Hesse, Benjamin Hinrichs for helpful comments. 

\section{Statements}

%CONFLICT OF INTEREST STATEMENT:  
The authors have no relevant financial or non-financial interests to disclose.
%FUNDING STATEMENT: 
No funding was received to assist with the preparation of this manuscript.
%DATA AVAILABILITY STATEMENT: 
Data sharing is not applicable to this article as no datasets were generated or analysed during the current study. 

\appendix
%\input{ResolventenGleichungv2}
%\input{ProofofAnaLemmav2}
%\input{A_Appendixv2}
%\input{Asupreumestimatev3}

%Befehl für BibLatex
\bibliography{Quellen}
\bibliographystyle{plain}
%\printbibliography

\end{document}